\documentclass[showpacs,12pt,a4paper]{article}  

\usepackage{geometry}
\usepackage{cite}
\usepackage{amsmath,amssymb}
\usepackage{latexsym,amscd,amsthm,amsxtra,graphicx,mathrsfs,amsfonts}  
\usepackage{verbatim}

\usepackage{dcolumn}
\usepackage[dvipdfm]{hyperref}
\hypersetup{CJKbookmarks=true}   

\usepackage{color}

\usepackage{multirow}
\usepackage{extarrows}
\usepackage{empheq,mdframed,lipsum}
\usepackage{diagbox}

\usepackage{caption,epsfig,subfig}

\usepackage{dsfont}
\usepackage{array}
\usepackage{booktabs}

\usepackage{xpatch}
\usepackage{blindtext}

\usepackage{authblk}
\usepackage{setspace}
\usepackage{lscape}

\usepackage{bm}
\usepackage{cases}

\usepackage{simplewick}   
\usepackage{pdfpages}

\newcommand{\be}{\begin{equation}}
\newcommand{\ee}{\end{equation}}
\newcommand{\ba}{\begin{eqnarray}}
\newcommand{\ea}{\end{eqnarray}}
\newcommand{\bea}{\begin{equation}\left\{ \begin{aligned}}
\newcommand{\eea}{ \end{aligned} \right. \end{equation}}
\newcommand{\bc}{\begin{numcases}  }
\newcommand{\ec}{\end{numcases}   }
\newcommand{\non}{\nonumber}

\newcommand{\ben}{\begin{equation*}}
\newcommand{\een}{\end{equation*}}
\newcommand{\ban}{\begin{eqnarray*} }
\newcommand{\ean}{\end{eqnarray*}}

\allowdisplaybreaks[2]

\begin{document}

\title{Field Theory with Fourth-order Differential Equations}

\author{
Rui-Cheng LI   \thanks{rui-chengli@163.com}\\
{\small\it  School of Physical Sciences, University of Chinese Academy of Sciences, Beijing 100049, China}
}

\date{}

\maketitle

\begin{abstract}
We introduce a new class of higgs type complex-valued scalar fields $U$ with Feynman propagator $\sim 1/p^4$
and consider the matching to the traditional fields with propagator $\sim 1/p^2$ in the viewpoint of effective potentials at tree level.
With some particular postulations on the convergence and the causality,
there are a wealth of potential forms generated by the fields $U$, such as
the linear, logarithmic, and Coulomb potentials,
which might serve as sources of effects such as
the confinement, dark energy, dark matter, electromagnetism and gravitation.
Moreover, in some limit cases, we get some deductions, such as:
a nonlinear Klein-Gordon equation,
a linear QED,
a mass spectrum with generation structure and a seesaw mechanism on gauge symmetry and flavor symmetry;
and,
the propagator $\sim 1/p^4$ would
provide a possible way
to construct a renormalizable gravitation theory and to solve the non-perturbative problems.
\\
\\
\end{abstract}





\thispagestyle{empty}

\newpage
\setcounter{page}{1}
\setcounter{tocdepth}{3}
\pagenumbering{roman}
\tableofcontents

\newpage
\setcounter{tocdepth}{3}
\pagenumbering{arabic}

\newpage

\section{Introduction}\label{introduction}

As a very successful theory, the gauge field theory with the gauge invariance principle could be used to solve a huge part of questions for people.
Certainly, there are some challenges to the gauge theory: the extension for methods of application such as the ones for non-perturbative problems, the extension for new phenomenons such as the ones for new particles and new interactions, with an inevitable old topic about the unification and renormalization.

It's just the linear potential from the non-perturbative results in lattice gauge theory\cite{HuangCY-book} that motivated us to consider a fourth order differential equations. The motivation chain is:
on the level of effective theories, we want to know what would be new in a theory which can generate a linear potential;
mathematically, a straightforward way to construct linear potential would be an introduction to Feynman propagator form of $1/p^4$, related to the higher order differential equations;
and, in the viewpoint of the superficial degree of divergence, the new Feynman propagator $\sim 1/p^4$ associated the higher order differential equations might provide us a construction to a renormalizable gravitation.
So, it would be significant to investigate models in the higher order differential equations formalism, in combining with the treatment on puzzles on the
redundant unphysical degrees of freedom (d.o.f).
For simplicity,
we will concentrate our studies on the pro forma feasibility of the model in a view of effective potentials at tree level.

The remainder of this paper is organized as follows:
Sect. 2 is for the Lagrangian construction for a linear potential;
Sect. 3 is for the kinetics and the propagators from the Lagrangian;
Sect. 4 is for the effective potentials generated from the Lagrangian, especially for the linear, Coulomb and gravitational potentials;
Sect. 5 is for some interesting deductions uniquely occurring in our theory for some limit cases;
Sect. 6 is for interpreting the causality in our theory;
and Sect. 7, the final section, is for our conclusions.

\section{Lagrangian for linear potential}\label{section-Lagrangian}

\subsection{Framework: effective potentials at tree-level}\label{formalism}

We can get the classic non-relativistic (NR) potential forms from the amplitudes of the tree-level  2$\rightarrow$ 2  scattering process for a perturbative theory, within the Born-approximation formalism, for instance, we can take\cite{Peskin}
\be
{\mbox{(vertex)}}_1\,\,\otimes \,\,\mbox{(inner-line propagator)}\,\,\otimes{\mbox{(vertex)}}_2 \Leftrightarrow {\mathcal V}
\ee
where the l.h.s is a part of the amplitude for a tree-level Feynman diagram, and the r.h.s is the classic potential. So, conversely, we can build theories for potentials with a definite form through the tree-level-correspondence, provided that the theories are perturbatively computable. For example, if there were neither momentums nor coordinates in the Feynman rules of vertices, we would extract different potentials with different inner-line propagators, such as:
\ba
\mbox{linear potential } &\leftrightarrow&   \frac{1}{p^4}\,,\nonumber\\
\mbox{Coulomb potential } &\leftrightarrow& \frac{1}{p^2} \,,\nonumber\\
\mbox{short-distance potential } &\leftrightarrow& \frac{1}{p^\alpha} \,,\mbox{with}\,\,\-\infty<\alpha<2.
\ea

\subsection{Lagrangian}\label{subsection-Lagrangian}

Firstly, we take a complex-valued scalar field $U$, a Dirac field $\psi$ (and $\bar{\psi}$) as the physical field degree of freedom(d.o.f)
\footnote{Discussions for the vector field $U^{\mu}$, the tensor field $U^{\mu\nu}$, and the massive $\{U, U^{\mu}\}$ have also been finished by the author, see Ref. \cite{LiRC-P4-theory}.},
which have the transformation law under a {\bf global} $U(1)$ group element $V$
\footnote{If we define $U=U_1 +i U_2$, then the global $U(1)$ symmetry, or a global $SO(2)$ symmetry, is defined between $U_1$ and $U_2$. }
as
\be
U\rightarrow VUV^{-1} =U \,,\, \psi \rightarrow V\psi\,,\bar{\psi} \rightarrow \bar{\psi} V^{-1}\,. \label{dof-trans-law}
\ee

Secondly,
in the method mentioned in Section \ref{formalism},
for a theory
with a propagator form $\thicksim\frac{1}{p^4}$ for $U$,
we write the Lagrangian of $\{U,\psi,\bar{\psi}\}$ as
\be
{\mathcal L}={\mathcal L}_{U}+{\mathcal L}_{\psi}+{\mathcal L}_{I},
\label{L-new}
\ee
where the term
\ba
{\mathcal L}_{U}&=&  Tr\left\{
- \partial^{\mu}\partial^{\nu}U^{\dag} \partial_{\mu} \partial_{\nu}U  - \Lambda_{U}^4 [(U+U^{\dag})+i (U-U^{\dag})]\right.  \non\\
&& \left.
+  m_{U}^4 U^{\dag}U
- \lambda_{U} \Lambda_{U}^4 U^{\dag}U U^{\dag}U
\right\}
\label{L-U-free}
\ea
is purely of the complex-valued scalar field $U$,
the term
\be
{\mathcal L}_{\psi}= \bar{\psi}(i\partial\!\!\!/ -m_{\psi})\psi
\ee
is purely of the matter field $\psi$,
and the term
\ba
{\mathcal L}_{I} &=&
  - \alpha \Lambda  Q_\alpha\,  \bar{\psi} \left\{[(U + U^{\dag})+i(U - U^{\dag})] \right\}  \psi  \nonumber\\
&&- \beta  Q_\beta\,   \bar{\psi} \left\{\sigma_{\mu\nu} \partial^\nu [( U + U^{\dag})+i( U - U^{\dag})] \right\}   \gamma^\mu \psi \nonumber\\
&& - \xi   \frac{1}{M} Q_\xi    \bar{\psi}  \left\{\sigma_{\mu\nu}  \partial^{\mu}[( U + U^{\dag})+i( U - U^{\dag})] \right\}   (i \overleftrightarrow{\partial}^{\nu}\psi)
\label{L-I}
\ea
is the invariant interaction term of $\psi$ coupled to $U$ under the transformations in (\ref{dof-trans-law}) and the Lorentz transformation.
The application of $\sigma^{\mu\nu}$ in the $\beta$ term is to ensure a real-valued effective coupling in the Feynman rule language, by recalling the reduction of
$ - i \sigma_{\nu\mu}    q^{\mu} \rightarrow    q_{\nu} $.\\

Thirdly, we give some {\bf postulations} as the illustrations of the variables in the Lagrangian of (\ref{L-new}) as below.\\

\indent 1. $\Lambda_{U}$ is a constant of the dimension of mass, $m_{U}$ is the mass of field $U$, and $\lambda_{U}$ is a dimensionless constant; $m_{\psi}$ is the mass of field $\psi$.\\

\indent 2. For the real-valued coefficients, there is $\alpha,\beta,\xi>0$. Particularly, if there is $\alpha=\beta=\xi$, there is a kind of symmetry between the intrinsic charges and the momentums of the matter fields $\psi$, which seems like a kind of realization of the supersymmetry.\\

\indent 3. For the parameters $\Lambda$ and $M$, referring to Wilson's scheme for renormalization, for the interaction Lagrangian terms we can propose the  postulations as: \\
\noindent (i)
each $U$ (not $\partial U$) is tied with one
infrared (I.R.)
energy scale   $ \Lambda $;\\
\noindent (ii)
all the terms with higher-dimensional ($D > 4$) are suppressed by a
ultraviolet (U.V.) energy scale $M$.

For example, if we plan to construct a QED, a QCD or a gravitation theory with the $U$ field, then the variable $\Lambda$ and $M$ might be respectively set as
\ba
&&\Lambda= \mu_{IR} \simeq  \{0,\Lambda_{QCD},0\},
M=\mu_{UV}  \simeq \{\mu_{EW}, \mu_{GUT},\mu_{Plank}\},
\label{energy-scale-1}
\ea
where $\mu_{IR}$ is the I.R. boundary energy scales, i.e., $\{$ value $\simeq0$, the QCD scale $\Lambda_{QCD} \simeq 200\, MeV$, value $\simeq0\}$,
and $M$ is the U.V. boundary for the theory,
i.e., $\{$the electroweak (EW) scale $\mu_{EW}\thicksim 246\,GeV$, the grand unification theory (GUT) scale $\mu_{GUT}$, the Plank scale $\mu_{Plank} \}$,
for a QED,a QCD and a gravitation theory, respectively.\\

\indent 4.  The variables $Q_{\{ \alpha,\beta,\xi\}}$ can be seemed as a kind of reconstructed charges (RC),
and they are defined as
\ba
Q_{\{ \alpha,\beta,\xi\}}  \equiv Y {\mathcal Q}_{\{ \alpha,\beta,\xi\}}, \, Y = \pm 1,\label{Q=YQ}
\ea
where
$Y$ is the generator of the global $U(1)$ group with eigenvalues $\pm 1$,
and ${\mathcal Q}_{\alpha}$ is a generator of some other global group (such as the electromagnetic $U(1)$ group) corresponding to the current ${\mathcal J}_\alpha$,
with the definitions
\ba
{\mathcal Q}_{\alpha} \equiv 1,\,Y;&&{\mathcal J}_\alpha  \equiv  \bar{\psi} \psi;\,\label{Qalpha}\\
{\mathcal Q}_{\beta} \equiv T_{\mathcal Q};&&{\mathcal J}_\beta \equiv  \bar{\psi}   \gamma^\mu \psi;\, \label{Qbeta}\\
{\mathcal Q}_{\xi} \equiv Y;&&{\mathcal J}_\xi \equiv \bar{\psi}i \overleftrightarrow{\partial\!\!\!/}  \psi ,\, \label{Qxi}
\ea
where  ${\mathcal Q}_{\alpha}=1$ for neutral $U$ (e.g. $U$ for mediating a QED theory), ${\mathcal Q}_{\alpha}=Y$ for charged $U$ (e.g. $U$ for  mediating a QCD theory);
$T_{\mathcal Q} \equiv T_{QED},T^a_{QCD},...$, is either the generator of the QED $U(1)$ group for constructing a QED theory with $U$, or one of the the generator of QCD $SU(3)$ group  for constructing a QCD theory with $U$, etc.

Furthermore, if we define a kind of effective media field as
\ba
({\mathcal A}_I)_\alpha &\equiv&-\alpha\Lambda  {\mathcal Q}_{\alpha} \cdot [(U + U^{\dag})+i(U - U^{\dag})],\, \\
\left[({\mathcal A}_I)_\beta\right]_{\mu}  &\equiv& -\beta {\mathcal Q}_{\alpha}\cdot\sigma_{\mu\nu}  \partial^\nu [(U + U^{\dag})+i(U - U^{\dag})],\,\label{effective-media-beta}
\ea
then the interaction Lagrangian terms in (\ref{L-I}) can be expressed as
\ba
{\mathcal L}_{I} \equiv {\mathcal L}^{RC} = ({\mathcal A}_I) \cdot {\mathcal J} \cdot Y.
\ea

\indent 5.  How to determine the value of $Y$ and ${\mathcal Q}_{\alpha}$? Here we define: if the momentum of $U$ flows ``in" to the $\bar{\psi}U\psi$ vertex, then the charge at this vertex is $Y=+1$, motivated by an imagination that the effective mass of $\psi$ would become bigger by ``eating" a nonzero vacuum expectation value $\langle U\rangle$; on the contrary, if the momentum of $U$ flows ``out" of the $\bar{\psi}U\psi$ vertex, then the charge at this vertex is $Y=-1$.
Similarly for the ${\mathcal Q}_{\alpha}$, ${\mathcal Q}_{\beta}$ and ${\mathcal Q}_{\xi}$, e.g.:\\
\indent (i) for ${\mathcal Q}_{\alpha}$: in the case of a charged $U$ for a QCD theory, in every physically allowed process, if the ${\mathcal Q}_{\alpha}$ charge of $U$ flows ``in" to the $\bar{\psi}U\psi$ vertex, then the ${\mathcal Q}_{\alpha}$  charge variation for the ``current" ${\mathcal J}_\alpha  \equiv  \bar{\psi}^i \psi^j$ (with $i,j$ the color indices) at this vertex is ${\mathcal Q}_{\alpha}=+1$, the same as the value of $Y$; on the contrary, if the ${\mathcal Q}_{\alpha}$ charge of $U$ flows ``out" of the $\bar{\psi}U\psi$ vertex, then the ${\mathcal Q}_{\alpha}$ charge variation for the ``current" ${\mathcal J}_\alpha  \equiv  \bar{\psi}^i \psi^j$ at this vertex is ${\mathcal Q}_{\alpha}=-1$, the same as the value of $Y$; in the case of a neutral $U$ for a QED theory, the ${\mathcal Q}_{\alpha}$ charge variation for the ``current" ${\mathcal J}_\alpha  \equiv  \bar{\psi} \psi$ at both vertices are defined to be always $1$;\\
\indent (ii) for ${\mathcal Q}_{\beta}$: even in the case of a neutral $U$ for a QED theory, the ${\mathcal Q}_{\beta}$ charge variation for the ``electromagnetic current" ${\mathcal J}_\beta  \equiv  \bar{\psi}\gamma^\mu \psi$ is not $1$, but to be the QED ``charge" $T_{QED}\equiv   {\mathcal Q}^{QED}$;\\
\indent (iii) for ${\mathcal Q}_{\xi}$: even in the case of a neutral $U$ for a gravitation theory, the ${\mathcal Q}_{\xi}$ charge variation for the  ``momentum current" ${\mathcal J}_\xi  \equiv  \bar{\psi}i \overleftrightarrow{\partial\!\!\!/}  \psi$ is not $1$, but to be $Y$; etc.\\

\indent 6. To ensure the renormalizability, we need an extra
{\bf postulation}: all divergences
can be removed by introducing cutoff for the amplitudes or the phase-space parameters.
More detail have been discussed in Ref. \cite{LiRC-P4-theory}.

\subsection{On the $(\partial\partial U)^2$ term for kinetics term}\label{ddU-in-LK}

The traditional kinetic term
\be
(\partial U )^2 \,\,\mbox{and} \,\,U^{\dag}\partial\partial U  \nonumber
\ee
will not appear in our model,
which is like the case that
a term $U^{\dag}\partial U $
will not appear in the kinetic term of a Klein-Gordon field; this might be related to a kind of generalized ``charge" symmetry. 
So, our theory with high-order derivatives is different from the ones discussed by Ostrogradski (or a quantized version by Pais, Uhlenbeck)\cite{Pais-Uhlenbeck}  or the so-called $f(R)$ theories discussed in general relativity formalism\cite{JPHsu}. \\

For convenience, we would call the model for $U$ defined with the $(\partial\partial U)^2$ term for kinetics term
as a ``{\bf P4 type}", and the traditional model for $U$ defined with  the $(\partial U)^2$ term for kinetics term
as a ``{\bf P2 type}". \\

It might be helpful for us to more easily understand the double partial term $(\partial\partial U)^2$ for the kinetics term, if
we understand our $U$ field as a classic continuum medium field.
For the detail, for the continuum medium field $\phi$ we have the continuity equation 
\ba
\partial_{\mu}\partial_{\nu}T^{\mu\nu}  =  0,  \label{continuity-equation}
\ea
with the energy-momentum tensor defined as
\ba
T^{\mu\nu} &=& (\rho+p)u^{\mu}u^{\nu}+pg^{\mu\nu}\nonumber\\
&=&\frac{\partial {\mathcal L}}{\partial(\partial_{\nu}\phi_{\alpha} )}\partial^{\mu}\phi_{\alpha} -g^{\mu\nu} {\mathcal L}\nonumber\\
&=&( \partial_{\mu} \phi^{\dag} \partial_{\nu}\phi +  \partial_{\nu} \phi^{\dag} \partial_{\mu}\phi)
 - g^{\mu\nu} ( \partial_{\alpha} \phi^{\dag} \partial^{\alpha}\phi -  m^2  \phi^{\dag} \phi)\nonumber\\
&=& \phi^{\dag} \left[\overleftarrow{\partial}_{\mu} \partial_{\nu}\phi + \overleftarrow{\partial}_{\nu} \partial_{\mu}
-  g^{\mu\nu} ( \overleftarrow{\partial}_{\alpha} \partial^{\alpha}  -     m^2)\right]\phi  \nonumber\\
&=&  \phi^{\dag} \left[(i\partial_{\mu} i\partial_{\nu}\phi + i \partial_{\nu}i \partial_{\mu})
-  g^{\mu\nu} (i\partial_{\alpha} i\partial^{\alpha}  -  m^2    )\right]\phi  . \label{energy-momentum-tensor}
\ea
Formally, to fully describe a field $\phi$, one might need the $\partial\partial\cdot\partial\partial$ operator acting on the field.

Moreover, we can write the E.O.M in another form,
\be
\hat{p}^4 U(x)= [\hat{p}^2  \Phi(x)]^2=[\hat{p}^2 \widetilde{\Phi}(x)] \cdot [\hat{p}^2\Phi(x)]   \,,
\label{equation-square-0}
\ee
with the correspondence for $\widetilde{\Phi} $ to $ \Phi$ here is just like a generalized version of the case that the anti-particles $\bar{\psi}$ associated with the particles $\psi$, which also arised from the treatment that the Dirac equation was formally from the square root of the Klein-Gordon equation. Besides, we can see, if the E.O.M is not the form $\hat{p}^4 U=m^4 U$, then that might break a generalized ``charge" symmetry between $\Phi$ and $\widetilde{\Phi}$.
We can denote that as
\ba
  \Phi(x)\thicksim \langle \bar{\psi}\psi \rangle
\Rightarrow &&  \mbox{K-G eq. } = \left[\mbox{ Dirac eq. }\right]^2   \,,\label{equation-square-1}\\
  U(x)\thicksim  \langle \widetilde{\Phi} \Phi \rangle
\Rightarrow && \mbox{U-eq. } = \left[\mbox{ K-G eq. }\right]^2    \,. \label{equation-square-2}
\ea
Then we can have the new E.O.M
\be
\hat{p}^2 \Phi  = m_{U}^2 \Phi \Rightarrow \Phi =c_1 e^{ip\cdot x}+c_2 e^{-ip\cdot x}   \label{field-solution-of-plane-wave}
\ee
for the ordinary physical d.o.f, and
\ba
&&\hat{p}^2 \widetilde{\Phi}  =  -m_{U}^2 \widetilde{\Phi} ,\,\,\,\,\mbox{(tachyon/higgs)} \label{EOM-tachyon}\\
&& -\hat{p}^2 \widetilde{\Phi}  =   m_{U}^2 \widetilde{\Phi},\,\,\,\,\,\,\,\,\,\mbox{(phantom)} \label{EOM-phantom}\\
\Rightarrow&&
\Phi = d_1 e^{p\cdot x}+d_2 e^{-p\cdot x}
\xrightarrow{\scriptsize \mbox{omit divergent terms}} \non\\
&& \quad \rightarrow
d_1 \left[e^{p \cdot x }\cdot \theta(-p \cdot x) +e^{-p\cdot x}\cdot\theta(p \cdot x)\right]
,\, (p\cdot x \neq 0),    \label{field-solution-of-instanton}
\ea
for the so-called unphysical d.o.f (with the $\theta$ function being the step function): the tachyons in (\ref{EOM-tachyon}), with an imaginary number valued mass\cite{tachyon};
and the phantoms in (\ref{EOM-phantom}), with a negative kinetic energy\cite{phantom},
respetively.
The sign of the action corresponding to the E.O.Ms in (\ref{EOM-tachyon}) and (\ref{EOM-phantom}) are different,
which is not negligible\cite{Weinberg}.

Although there exist unphysical and acausal solutions in addition to the two physical d.o.f
for differential equations with orders higher than 2 in the classic mechanics case,
\footnote{The acausality discussed in Ref. \cite{Itzykson-Zuber} only occurs in the classic mechanics case and can be removed in the formalism of quantum mechanics through the uncertainty principle by treating all the observable variables as operators.}
we can avoid this trouble  by treating these solutions as effects of hidden unphysical new d.o.f (which are existent but can't be directly measured for some reasons) beyond the standard model (SM) in particle physics; this is to discussed in the following Section \ref{higgs-U}. We will revisit this topic in Section \ref{section-causality}, and we want to propose that unphysical d.o.f does not necessarily mean acausality.

\subsection{$U$ is a kind of higgs-type field!}\label{higgs-U}

The self-interaction potential of field $U$ is
\ba
V(U)\equiv -  m_{U}^4 U^{\dag}U
+ \lambda_{U} \Lambda_{U}^4 U^{\dag}U U^{\dag}U, \label{potential-U}
\ea
so, according to the minus sign in the mass term, $U$ is a kind of higgs-type field. And, for convenience, in all this article for allowed cases we set
\ba
\langle U \rangle =1.   \label{set-U-VEV=1}
\ea
But we should remind ourselves that $\langle U \rangle $ could be very large even when the energy scale is very low.

\begin{figure}
\noindent \begin{minipage}[t]{0.45\linewidth}
\centering
\includegraphics[scale=0.29]{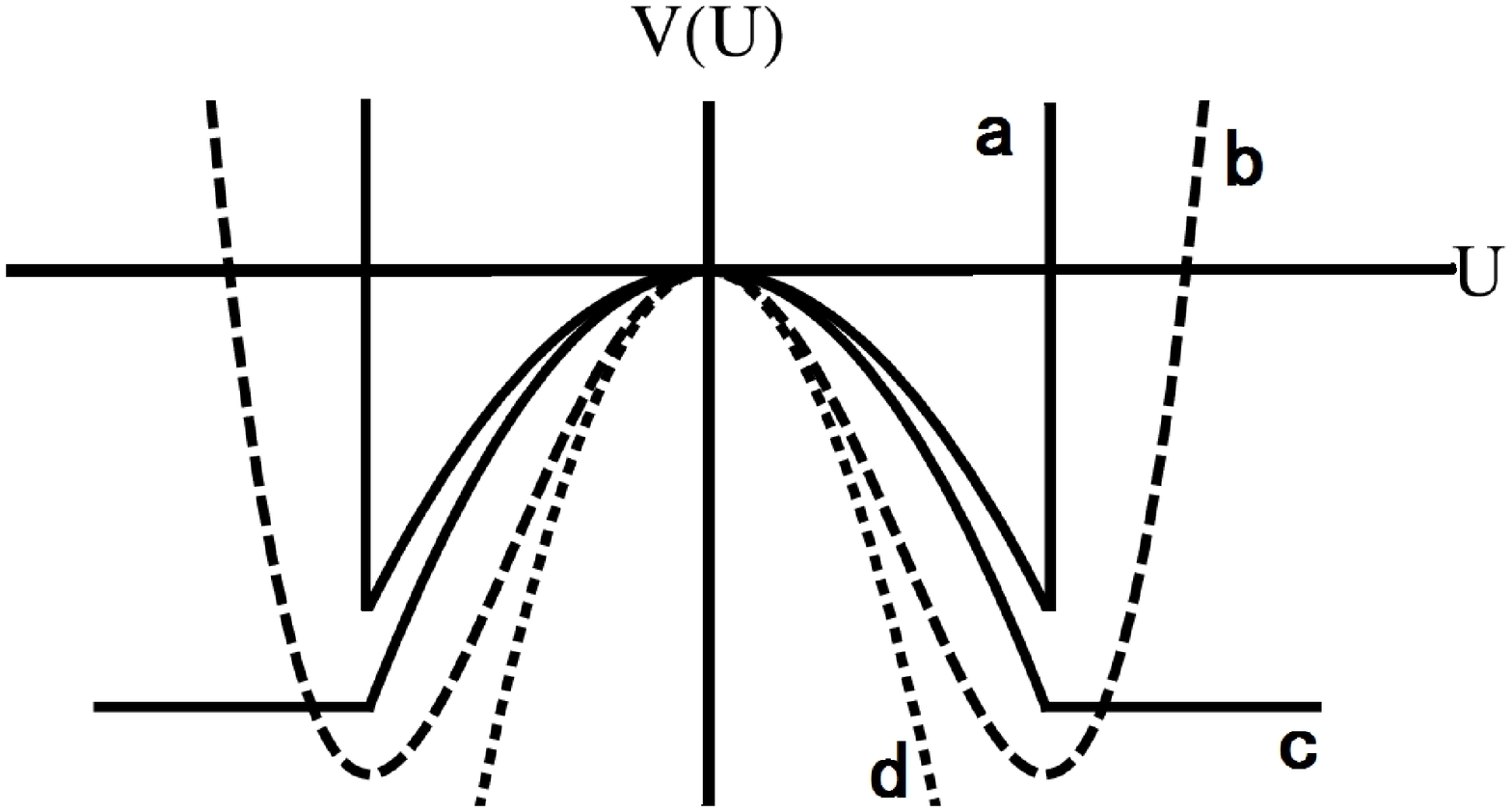}
\centerline{(1)}
\end{minipage}
\quad\quad
\noindent \begin{minipage}[t]{0.45\linewidth}
\centering
\includegraphics[scale=0.25]{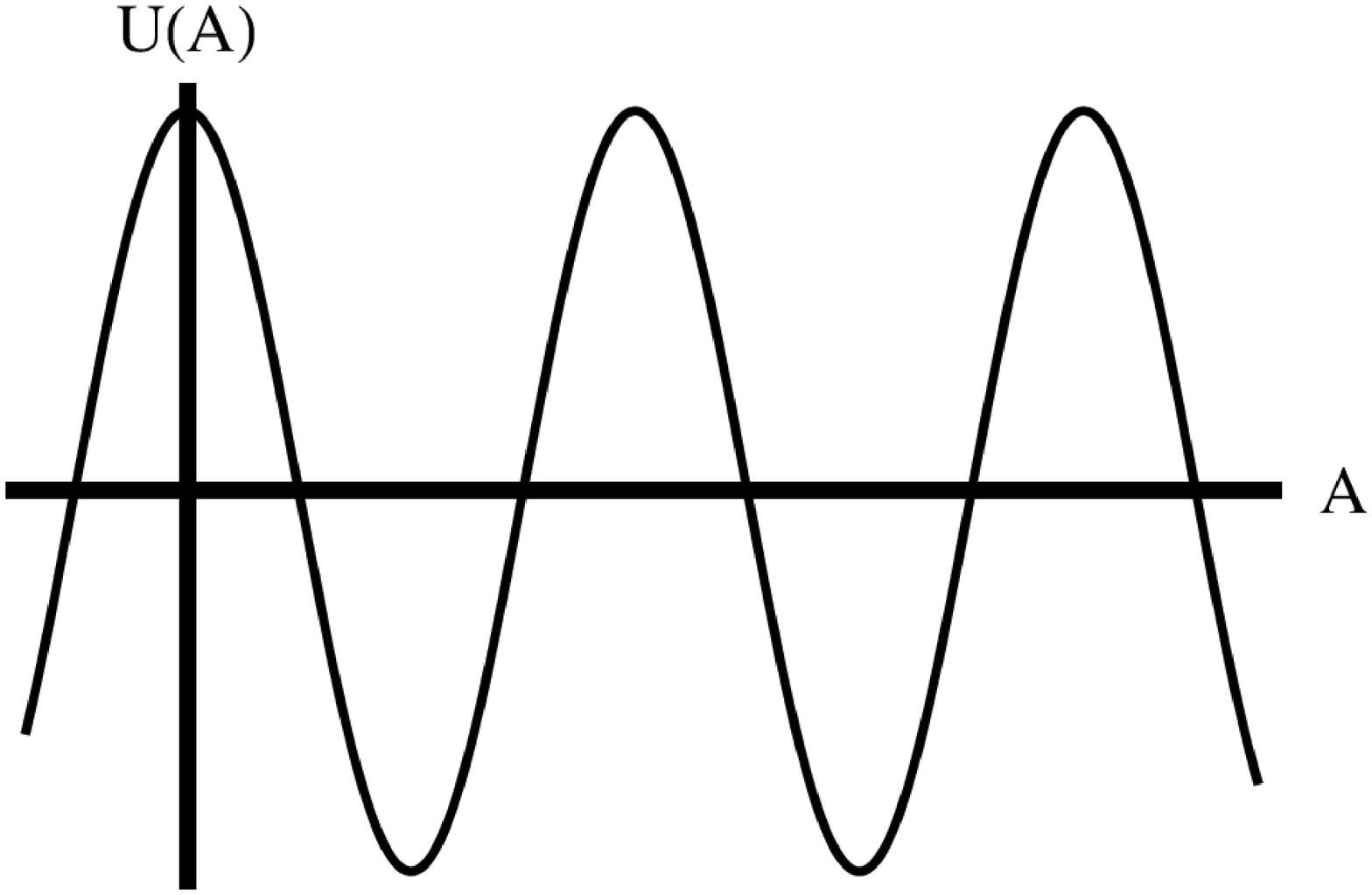}
\centerline{(2)}
\end{minipage}
\caption{Self-interaction potentials for the field $U$ and $A$.}
\label{U-potentails}
\end{figure}

For a higgs field $U$ with a potential form in (\ref{potential-U}) plotted as the line-``b" in Fig.-\ref{U-potentails}-(1), besides of the angular component $U_\theta$ as the conventional field (the Goldstone boson), there is also a radial-direction component $0\leq U_r \leq +\infty$.
Here, the most important point is, how to understand the $U_r$?

For a potential $V(U)$ of the form as the line-``a" in Fig.-\ref{U-potentails}-(1), which is defined only for $0\leq |U|\leq 1$ rather than for all the $|U|<\infty$ field configurations,
we can not only treat the radial-direction component $U_r$ as a stable (physical) fluctuation around the stable vacuum $|U|=1$ (minimum of the potential $V(U)$),
but also treat $U_r$ as a  $0\leq U_r \leq 1$ oscillating around the point $|U|=0$ maintained by the rebound from the potential barrier.
Similarly, for a potential $V(U)$ of the form as the line-``b" in Fig.-\ref{U-potentails}-(1),
we can also understand
the radial-direction component $0\leq U_r \leq +\infty$
in two viewpoints:
$ U_r$ is a stable (physical) field d.o.f $U_r^{higgs}$ oscillating around the stable vacuum $|U|=1$ (minimum of the potential $V(U)$), which could be seemed as the ``traditional" P2 type excitation of ``higgs particle"; or, $ U_r^{P4}$ is an unstable (unphysical) field d.o.f  oscillating around the unstable vacuum $|U|=0$ (local maximum of the potential $V(U)$), which would ``decay/collapse" as what would happen in the more extreme two cases plotted as the line-``c" or line-``d" in Fig.-\ref{U-potentails}-(1)).

However,
we will just take the unstable (unphysical) $ U_r^{P4}$ d.o.f as the real component in our ``untraditional" P4 type $U$ field, with the purpose to design the $U$ field to differ from the ``traditional" P2 type field.
Thus, from now on, we need not give too many query to the sign of the mass term in (\ref{L-U-free}) any more.
As discussed in Section \ref{ddU-in-LK}, we can say:
$U$ is a kind of higgs-type field,
and $U$ does have a nonzero VEV,
however, the $U$ field with E.O.M. $\hat{p}^4 U = m^4 U$ is really designed to be
neither a traditional higgs field  with E.O.M. $\hat{p}^2 U =-m^2 U$ nor a phantom with E.O.M. $-\hat{p}^2 U = m^2 U$, see (\ref{EOM-tachyon}).

In a word, it should be emphasized that the choice for the sign of the mass term is very important and crucial for our following work.\\

\noindent 

\section{The kinetics}\label{section-kinetics}

\subsection{The equation of motion of the $U$ field}\label{section-EOM}

By the Euler-Lagrange equation \cite{ChengTP-book}
\be
\frac{\partial {\mathcal L}_{U} }{\partial  U } -  \partial_{\mu}\frac{\partial {\mathcal L}_{U} }{\partial(\partial_{\mu} U )}
 +  \partial_{\mu}\partial_{\nu} \frac{\partial {\mathcal L}_{U} }{\partial(\partial_{\mu}\partial_{\nu}  U )} =0\,,
\label{Euler-Lagrange-equation}
\ee
from (\ref{L-U-free}) we can get the equation of motion(E.O.M) of free field U,
\ba
&-\partial^{\mu}\partial^{\nu}\partial_{\mu}\partial_{\nu}U & =-m_{U}^4 U +\Lambda_{U}^4 \\
\Leftrightarrow &  -\hat{p}^4 U &=-m_{U}^4 U  +\Lambda_{U}^4   \,,\,\hat{p}^{\mu}=i\partial^{\mu}\,,
\label{EOM-U}
\ea
and the dynamical E.O.M for $U$,
as
\be
-\partial^4 U  =    - m_{U}^4  U    + \Lambda_{U}^4    +  \alpha Q  \Lambda\, \bar{\psi}  \psi +... \,.
\label{EOM-U-scalar}
\ee

\subsection{The propagator}\label{canonic-commutator-and-propagator}

By inserting the ``correlation function", i.e., one version of the definitions of propagator of $U$,
\ba
D_F(x-y)&\equiv& \langle \Omega| \hat{\rm T} U(x)U(y)|\Omega\rangle \nonumber\\
&=& \theta(x^0 - y^0)\langle \Omega| U(x)U(y) |\Omega\rangle+ \theta(y^0 - x^0)\langle \Omega| U(y)U(x) |\Omega\rangle   \label{define-correlation-function}
\ea
into the E.O.M, where $\hat{\rm T}$ is the time-ordering operator, $|\Omega\rangle$ is the vacuum state, we can verify
\ba
&&-(\partial^4\,-m^4\,)_{x}D_F(x-y)\non\\
&\equiv& (\partial^4\,-m^4\,)_{x}\langle \Omega| \hat{\rm T} U(x)U(y)|\Omega\rangle
= + i\delta^{(4)}(x-y) \,.
\label{propagater-U-scalar-checksign}
\ea
That means, $D_F(x-y)$ is really the ``Green function", i.e., the other version of the definitions of propagator of $U$. \\

By setting $\Lambda_{U}=0$, from (\ref{propagater-U-scalar-checksign}) or its corresponding form in the momentum space
\ba
 -(p^4 - m_{U}^4  )\widetilde{D}_{F}(p)  =\,\, i \,, \label{EOM-U-scalar-momentum-space}
\ea
we can get the Feynman propagator for $m_U\neq 0$ case
in the momentum space, as
\be
\widetilde{D}_{F}(p) = \frac{-i }{p^4- m_{U}^4+i\epsilon} =\frac{-i}{(p^2+m_U^2 - i\epsilon)(p^2-m_U^2 + i\epsilon)},  \,\,
\mbox{($\Lambda_{U}=0$, $m_U\neq 0$)} ,
\label{propagator-U-scalar}
\ee
or, for $m_U = 0$ case
\be
D_F(U)=\frac{-i}{p^4+ i\epsilon} ,  \,\, \mbox{($\Lambda_{U}=0$, $m_U=0$)} .
\label{propagator-U-scalar-massless}
\ee

So, the minus sign before the $\hat{p}^4$ operator in the E.O.M (\ref{EOM-U},\ref{EOM-U-scalar},\ref{propagater-U-scalar-checksign},\ref{EOM-U-scalar-momentum-space}) is very crucial, which represents the sign of the mass term in Lagrangian, and, without this ``$-1$" factor, everything will be different! After all, the $U$ here isn't the traditional
scalar field, as we said in Section \ref{higgs-U}.
Besides,
the position and residue of a pole in the propagator is crucial for the calculation results of the amplitudes. In this work, we will only consider the $m_U=0$ case, and the $m_U \neq 0$ case has been discussed in Ref. \cite{LiRC-P4-theory}.

\section{Effective potentials}\label{section-potential}

\noindent\begin{figure}[!htbp]
\centering
\includegraphics[scale=0.7]{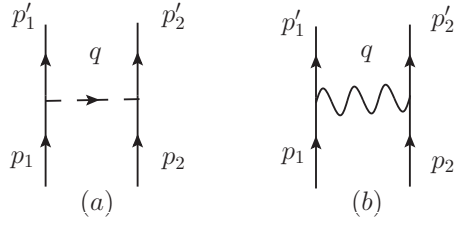}
\caption{The Feynman diagrams for the leading order tree level processes, with (a) mediated by a $U$ and (b) mediated by a photon $A^\mu$.}
\label{fig-tree-LO}
\end{figure}

At the beginning, we set the variables for the particles in the scattering processes shown in Fig. \ref{fig-tree-LO}, as below:
\ba
p_1=(m,{\bm p}_1),\, && p_2=(m,{\bm p}_2), \\
p'_1=(m,{\bm p}'_1),\,&& p'_2=(m,{\bm p}'_2)\,.
\ea
In the non-relativistic approximation, $q^0=0$ (which is also called on-shell approximation), we have the relations for kinetics variables as
\be
q = p_1-p'_1 \Rightarrow   q^2=(p_1-p'_1)^2  \xlongequal[q^0=0]{\mbox{\scriptsize(NR\,limit)}}
- |{\bm q}|^2
=-|{\bm p}_1- {\bm p}'_1|^2, \label{q0=0}
\ee
and
\be
\bar{u}^{s'}(p')              u^{s}(p)
=        2m\delta^{ss'},\,
\bar{u}^{s'}(p') \gamma^{\mu} u^{s}(p) \xlongequal{\mbox{\scriptsize(NR\,limit)}}
v^{\mu}  2m\delta^{ss'} \,.
\label{spinor-basis}
\ee

Besides, In the non-relativistic limit, we need not consider the identical particle effects, that is, we need not consider the $u$ channel of the Feynman diagrams in a scattering process now.

\subsection{Interaction I: coupled to intrinsic charges, Coulomb force?}\label{section-potential-Coulomb}

Now, for the interaction term
\ba
{\mathcal L}_{\alpha\beta} &=&
  - \alpha\Lambda  Q_\alpha\,\bar{\psi}  \left\{[(U + U^{\dag})+i(U - U^{\dag})] \right\}   \psi  \nonumber\\
&&- \beta  Q_\beta\,   \bar{\psi} \left\{\sigma_{\mu\nu} \partial^\nu [( U + U^{\dag})+i( U - U^{\dag})] \right\}   \gamma^\mu \psi ,
  \label{L-I-Coumlomb}
\ea
which was extracted from the total interaction Lagrangian ({\ref{L-I}}), by defining the couplings
\ba
\alpha_{1,2}\equiv \alpha (Q_\alpha)_{1,2} ,\,
\beta_{1,2}= \beta (Q_\beta)_{1,2} ,\label{alpha1-and-beta1-define}
\ea
and by using
$i \sigma_{\mu\nu} q^\nu =  - q_{\mu} $,
$\gamma^\mu \rightarrow v^\mu$ from (\ref{spinor-basis}),
we can write the corresponding amplitude for Fig. \ref{fig-tree-LO}-(a), as\footnote{For simplicity, here we can only consider the contributions from $U_1$, and, for the contributions from $U_2$, the result just need a double.}
\ba
i{\mathcal M}_{a}
&= &
\bar{u}^{s'} i[  -\alpha_1\Lambda - \beta_1  \sigma_{\mu\nu} (iq^\nu)  \gamma^\mu  ] u^{s}
\cdot \frac{-i}{q^4} \non\\
&&\cdot
\bar{u}^{r'} i[  -\alpha_2\Lambda - \beta_2  \sigma_{\alpha\beta} (-iq^\beta)    \cdot \delta_{\mu\alpha}  \gamma^\alpha   ]u^{r} \non\\
& =&
\bar{u}^{s'} i[  -\alpha_1\Lambda + \beta_1   q_\mu   \gamma^\mu  ] u^{s}
\cdot \frac{-i}{q^4} \cdot
\bar{u}^{r'} i[  -\alpha_2\Lambda - \beta_2   q_{\alpha} \cdot \delta_{\mu\alpha}  \gamma^\alpha   ]u^{r} \non\\
& =&
\bar{u}^{s'} [  -i\alpha_1 \Lambda] u^{s}
\cdot \frac{-i}{q^4} \cdot
\bar{u}^{r'} [  -i\alpha_2 \Lambda  ]u^{r}  \non\\
&&+
  [ \alpha_2\Lambda \beta_1 (q \cdot v_1)  -    \alpha_1\Lambda \beta_2 (q  \cdot  v_2)]
\cdot \frac{-i}{q^4} \cdot
\bar{u}^{s'} u^{s} \bar{u}^{r'}u^{r} \non\\
&&+
\bar{u}^{s'} [  i\beta_1     \gamma^\mu  ] u^{s}
\cdot \frac{-i   g_{\mu\alpha} }{q^2} \cdot
\bar{u}^{r'} [- i\beta_2   \gamma^\alpha   ]u^{r} ,\label{amp-alpha-beta}
\ea
where the indices in $\delta_{\mu\alpha}$ are not controlled by
the Einstein summation convention, and we have taken the replacement
\ba
q_\mu  q_{\alpha}  \delta_{\mu\alpha} \rightarrow   q^2   g^{\mu\alpha} .\label{qq-to-qsgmunu}
\ea
The use of the $\delta_{\mu\alpha}$ in (\ref{amp-alpha-beta}) could be understood by this reason: as there is only the single $U$ field exchanged in a $2\rightarrow 2$ scattering process,  once the $\mu$-component
of the momentum $q^{\mu}$ of $U$
is absorbed into one vertex in the Feynman diagram,
there must be the same $\mu$-component of the momentum $q^{\mu}$ is emitted out from the other vertex in the Feynman diagram!
Thus,
by treating $\partial^{\mu} U \sim \mathcal{A}^{\mu}$ as an P2 type effective media field as in (\ref{effective-media-beta}), for which there is a propagator with the form of $\frac{-i g^{\mu\alpha}}{q^2}$,
we can see, the last term in (\ref{amp-alpha-beta}) is just of the form of an amplitude for a scattering process corresponding to the Coulomb potential in QED, as shown in Fig.\ref{fig-tree-LO}-(b).

In the NR limit of $q^0=0$,
with the definitions
\ba
 {\bm q} \cdot{\bm v}_1 \equiv \lambda_1 |{\bm q}|,\,{\bm q} \cdot{\bm v}_2\equiv \lambda_2  |{\bm q}|,\,
\ea
and the approximation
$\gamma^\mu \gamma^\alpha g_{\mu\alpha} \rightarrow \gamma^0 \gamma^0 g_{00} =1 $,
we can continue to get
\ba
i{\mathcal M}_{a}
&= & -i \left\{
-  \frac{ \alpha_1 \alpha_2 \Lambda^2}{|{\bm q}|^4 }
+  \frac{(\alpha_2\beta_1 \lambda_1   -    \alpha_1\beta_2 \lambda_2 )\Lambda}{|{\bm q}|^3}
-  \frac{ \beta_1  \beta_2  }{|{\bm q}|^2}
\right\}
\cdot \bar{u}^{s'}u^{s} \cdot \bar{u}^{r'}u^{r} ,\,\, \label{tree-amp-a}
\ea
The amplitude $i{\mathcal M}$ should be compared with the Born approximation to the scattering amplitude in non-relativistic quantum mechanics, written in terms of the potential function $V({\bm x})$:\cite{Peskin} 
\ba
i{\mathcal M} \thicksim\, _{NR}\langle p'|iT|p \rangle _{NR} = -i \widetilde{V}({\bm q})(2\pi)\delta(E_{{\bm p}'}-E_{{\bm p}}),\,
{\bm q}={\bm p}-{\bm p}',\,\label{Born-approximation}
\ea
with
\ba
{\bm p} = \eta_2{\bm p}_1 - \eta_1{\bm p}_2,\,
{\bm p}' = \eta_2{\bm p}'_1 - \eta_1{\bm p}'_2,\,
\eta_{i}=\frac{m_i}{m_1+m_2},\, i=1,2   .
\ea
By dealing with the kinetics factors as $2m\delta^{ss'}\rightarrow \delta^{ss'}$ and $(2\pi)\delta(E_{{\bm p}'}-E_{{\bm p}})\rightarrow 1$, we can have
\ba
\widetilde{V}({\bm q})=
-  \frac{ \alpha_1 \alpha_2 \Lambda^2}{|{\bm q}|^4 }
+  \frac{(\alpha_2\beta_1 \lambda_1   -    \alpha_1\beta_2 \lambda_2 )\Lambda}{|{\bm q}|^3}
-  \frac{ \beta_1  \beta_2  }{|{\bm q}|^2} ,
\ea
and the inverse Fourier transformation
\be
V({\bm x})  =  {\mathcal F}^{-1}[\widetilde{V}({\bm q})] \,.
\ee
Then,
we can get the potential
\ba
V(r)  =
+ \frac{\alpha_1 \alpha_2 \Lambda^2}{8\pi} r
- \frac{(\alpha_2\beta_1 \lambda_1   -    \alpha_1\beta_2 \lambda_2 )\Lambda }{2\pi^2 } ({\rm log\,} \frac{r}{r_0}+\gamma_E -1)
+ \frac{-  \beta_1  \beta_2}{4\pi r } ,      \label{U-whole-potential-0}
\ea
with $r_0=1\,GeV^{-1}$ put by hand to balance the dimension, and $\gamma_E$ the Euler constant.
Moreover, by applying
(\ref{Q=YQ},\ref{Qalpha},\ref{Qbeta},
\ref{alpha1-and-beta1-define})
to get\footnote{It is very important to set $Y_1 = -1$ and $Y_2 = +1$ to match the Feynman rules used in Eq. (\ref{amp-alpha-beta})!}
\ba
&&\mbox{(for QED:)}\non\\
\alpha_1 \alpha_2
&=& \alpha^2  ( Q_\alpha)_{1} ( Q_\alpha)_{2}
= \alpha^2  ( Y \cdot 1 )_{1}  ( Y \cdot 1 )_{2}
= - \alpha^2  ,\label{alpha1-dot-alpha2}\\
- \beta_1 \beta_2
&=& - \beta^2 ( Q_\beta)_{1} ( Q_\beta)_{2}
= - \beta^2  ( Y {\mathcal Q}_{\beta} )_{1}  ( Y {\mathcal Q}_{\beta} )_{2}
\non\\
&=&
 - \beta^2 ( Y_{1} {\mathcal Q}^{QED}_{1}) ( Y_{2} {\mathcal Q}^{QED}_{2})
=   \beta^2     {\mathcal Q}^{QED}_{1} {\mathcal Q}^{QED}_{2} ,\label{beta1-dot-beta2}\\
\alpha_1\beta_2
&=& \alpha \beta  ( Q_\alpha)_{1} (Q_\beta)_{2}
= \alpha \beta  (Y \cdot 1)_{1} (Y  Q^{QED})_{2}
=- \alpha \beta    {\mathcal Q}^{QED}_{2}, \label{alpha1-dot-beta2}\\
\alpha_2\beta_1
&=& \alpha \beta  ( Q_\alpha)_{2} (Q_\beta)_{1}
=  \alpha \beta  ( Y \cdot 1 )_{2}  (Y  Q^{QED})_{1}
= - \alpha \beta     {\mathcal Q}^{QED}_{1},\label{alpha2-dot-beta1}
\ea
with ${\mathcal Q}^{QED}$ just the electric charge in QED,
and
\ba
&&\mbox{(for QCD:)}\non\\
\alpha_1 \alpha_2
&=& \alpha^2  ( Q_\alpha)_{1} ( Q_\alpha)_{2}
= \alpha^2  ( Y \cdot Y )_{1}  ( Y \cdot Y )_{2}
= \alpha^2  ,\label{alpha1-dot-alpha2-QCD}\\
- \beta_1 \beta_2
&=&
-\beta^2     {\mathcal Q}^{QCD}_{1} {\mathcal Q}^{QCD}_{2} ,\label{beta1-dot-beta2-QCD}\\
\alpha_1\beta_2
&=& \alpha \beta  (Y \cdot Y)_{1} (Y  Q^{QCD})_{2}
= \alpha \beta  (Y_{2}  Q^{QCD}_{2})
= \alpha \beta    {\mathcal Q}^{QCD}_{2}, \label{alpha2-dot-beta1-QCD}\\
\alpha_2\beta_1
&=&  \alpha \beta  ( Y \cdot Y )_{2}  (Y  Q^{QCD})_{1}
=  \alpha \beta    (Y_{1}  Q^{QCD}_{1})
= -\alpha \beta     {\mathcal Q}^{QCD}_{1},\label{alpha1-dot-beta2-QCD}
\ea
with ${\mathcal Q}^{QCD}$ just the color charge in QCD,
by combining with (\ref{alpha1-and-beta1-define},\ref{alpha1-dot-alpha2},\ref{beta1-dot-beta2},\ref{alpha1-dot-beta2},\ref{alpha2-dot-beta1}),
the potential in (\ref{U-whole-potential-0}) will become
\ba
V(r)_{QED} &=&
-  \frac{\alpha^2 \Lambda_{QED}^2}{8\pi}\cdot r
+ \frac{\alpha\beta \Lambda_{QED}(   {\mathcal Q}^{QED}_{1} \lambda_1  -  {\mathcal Q}^{QED}_{2} \lambda_2  ) }{2\pi^2 } \cdot  {\rm log\,} \frac{r}{r_0}  \non\\
&&+ \frac{\beta^2     {\mathcal Q}^{QED}_{1} {\mathcal Q}^{QED}_{2} }{4\pi r }       ,\label{U-whole-potential}\\
V(r)_{QCD} &=&
 \frac{\alpha^2 \Lambda_{QCD}^2}{8\pi}\cdot r
+ \frac{\alpha\beta \Lambda_{QCD}(   {\mathcal Q}^{QCD}_{1} \lambda_1  +  {\mathcal Q}^{QCD}_{2} \lambda_2  ) }{2\pi^2 } \cdot  {\rm log\,} \frac{r}{r_0}  \non\\
&&- \frac{\beta^2     {\mathcal Q}^{QCD}_{1} {\mathcal Q}^{QCD}_{2} }{4\pi r }       ,\label{U-whole-potential-QCD}
\ea
By recalling that we have performed the derivations in the formalism of a collision process in the center-of-mass frame,
that is to say, ${\bm p}_1=-{\bm p}_2$, by combining the on-shell approximation $|{\bm p}_1|=|{\bm p}'_1|$, we can indeed determine the relations
\ba
&& {\bm q} \cdot ({\bm p}_1 +   {\bm p}_2)  =   ( m_1 \lambda_1  +  m_2 \lambda_2  ) |{\bm q}| = 0  ,\,\non\\
&& {\bm q} \cdot ({\bm p}_1 -     {\bm p}_2)  =   ( m_1 \lambda_1  -  m_2 \lambda_2  ) |{\bm q}| >0  ,\,\non\\
&&( {\bm q} \cdot {\bm p}_1 )( {\bm q} \cdot {\bm p}_2) = m_1m_2   \lambda_1   \lambda_2  |{\bm q}|^2  < 0  .\label{lambda-sign}
\ea
So, for the case of $m_1=m_2$, we will have $ \lambda_1  -    \lambda_2    >0 $ and $\lambda_1 + \lambda_2 =0$.
As in (\ref{amp-alpha-beta}), we can see again,
the last term in (\ref{U-whole-potential}) or (\ref{U-whole-potential-QCD})
is coincidentally for the Coulomb interaction in QED or QCD, respectively!

Besides, there is a linear potential and a logarithmic potential in both (\ref{U-whole-potential}) and (\ref{U-whole-potential-QCD}).
In (\ref{U-whole-potential}), since the infrared energy scale boundary $\Lambda_{QED}$ for the QED is about zero, the linear potential and the logarithmic potential could be negligible;
however, in some cosmological experiments, the linear and the logarithmic term might give
corrections to the electromagnetic observables, such as: a spatial variation of the electromagnetic fine structure constant \cite{e-measurement},
or a kind of electromagnetic red-shift coupled with the gravitational red-shift.
In(\ref{U-whole-potential-QCD}), since the infrared energy scale boundary $\Lambda_{QCD}$ for the QCD is about $200\,MeV$, the linear potential could be significant to serve as the major part of the confinement in QCD, while the logarithmic potential could serve as a minor part of the confinement. 　

\subsection{Interaction II: coupled to momentum, gravitation?}\label{section-potential-gravity}

Now we consider the interaction terms,
\ba
{\mathcal L}_{\alpha\xi} &=&
- \alpha \Lambda  Q_\alpha\,  \bar{\psi} \left\{[(U + U^{\dag})+i(U - U^{\dag})] \right\}  \psi  \nonumber\\
&& - \xi   \frac{1}{M} Q_\xi    \bar{\psi}  \left\{\sigma_{\mu\nu}  \partial^{\mu}[( U + U^{\dag})+i( U - U^{\dag})] \right\}   (i \overleftrightarrow{\partial}^{\nu}\psi)  ,  \label{L-log}
\ea
which was extracted from the total interaction Lagrangian ({\ref{L-I}}).
By defining the couplings as in (\ref{alpha1-and-beta1-define})
\ba
\alpha_{1,2}\equiv \alpha  (Q_\alpha)_{1,2} ,\,
\xi   _{1,2}\equiv \xi     (Q_\xi   )_{1,2}
\ea
and by using
$      i \sigma_{\mu\nu}    q^\nu =  - q_{\mu} $,
$q_{\nu } q_{\beta } \delta_{\nu\beta}  \rightarrow   q^2   g_{\nu\beta}  $ as in (\ref{amp-alpha-beta},\ref{qq-to-qsgmunu}),
we can write the corresponding amplitude for Fig. \ref{fig-tree-LO}-(a), as
\ba
i{\mathcal M}
&=&\bar{u}^{s'}  i  \left\{ -\alpha_1 \Lambda -  \frac{\xi_1}{M}   \sigma_{\mu\nu} \cdot (i q^{\mu})\cdot \left[ i\cdot i (p_1+p'_1)^{\nu} \right]   \right\}    u^{s}
\cdot \frac{-i}{q^4} \nonumber\\
&&\cdot \bar{u}^{r'} i \left\{ -\alpha_2 \Lambda -  \frac{\xi_2}{M}   \sigma_{\alpha\beta} \cdot (-i q^{\alpha}) \cdot\delta_{\nu\beta}\cdot\left[i\cdot i(p_2+p'_2)^{\beta}  \right]       \right\}    u^{r}    \nonumber\\
&=& \frac{-i}{q^4} \cdot
\left\{
 - \alpha_1 \alpha_2 \Lambda^2
+ \frac{ \Lambda }{M}  q  \cdot \left[ \alpha_2 \xi_1 (p_1+p'_1)  - \alpha_1 \xi_2 (p_2+p'_2)  \right]    \right.\nonumber\\
&&\left.
+\frac{\xi_1 \xi_2}{M^2}    \cdot  q_{\nu } q_{\beta } \delta_{\nu\beta} \cdot   (p_1+p'_1)^{\nu}   (p_2+p'_2)^{\beta}
\right\}   \cdot \bar{u}^{s'} u^{s}\cdot \bar{u}^{r'}       u^{r}   \non\\
&=& -i \cdot \left\{
  -     \frac{\alpha_1  \alpha_2 \Lambda^2  }{ |{\bm q}|^4  }
-  \frac{ \Lambda}{M}\cdot \left[  \frac{ ( 2 \alpha_2 \xi_1  {\bm q}  \cdot {\bm p}_1  -2 \alpha_1 \xi_2   {\bm q}  \cdot {\bm p}_2    )}{|{\bm q}|^4} \right]\right.\nonumber\\
&&
+   \frac{ \Lambda}{M}\cdot \left[  \frac{(\alpha_1 \xi_2   +  \alpha_2 \xi_1 )   }{|{\bm q}|^2}  \right]
-  \frac{\xi_1\xi_2}{M^2}    \cdot  \frac{ 4  p_1\cdot   p_2  }{ |{\bm q}|^2} \nonumber\\
&&\left.
+  \frac{\xi_1\xi_2}{M^2}    \cdot   \frac{2  (  {\bm q}\cdot {\bm p}_1  -  {\bm q}\cdot {\bm p}_2   ) }{ |{\bm q}|^2}
-   \frac{\xi_1\xi_2}{M^2}
\right\} \cdot 2m\delta^{ss'} 2m\delta^{rr'}   ,\, |{\bm q}|>0.
\ea

In the non-relativistic limit, with ${\bm p}_{1,2} = m_{1,2} {\bm v}_{1,2}$,
and the definitions
\ba
 {\bm q} \cdot{\bm v}_1 \equiv \lambda_1 |{\bm q}|,\,{\bm q} \cdot{\bm v}_2\equiv \lambda_2  |{\bm q}|,\,
\ea
we can get
\ba
&& {\bm q} \cdot  {\bm p}_1   =     m_1 \lambda_1  |{\bm q}|, \quad
   {\bm q} \cdot  {\bm p}_2   =     m_2 \lambda_2  |{\bm q}|.
\ea
Thus the non-relativistic effective potential in the momentum space  will be
\ba
&&\widetilde{V}({\bm q})=  -\mathcal{M} \non\\
&=&
-  \frac{\alpha_1  \alpha_2 \Lambda^2  }{ |{\bm q}|^4  }
-  \frac{ \Lambda}{M}\cdot \left[  \frac{ ( 2 \alpha_2 \xi_1  m_1 \lambda_1     -2 \alpha_1 \xi_2   m_2 \lambda_2     )}{|{\bm q}|^3} \right] \nonumber\\
&&
+   \frac{ \Lambda}{M}\cdot \left[  \frac{(\alpha_1 \xi_2   +  \alpha_2 \xi_1 )   }{|{\bm q}|^2}  \right]
-  \frac{\xi_1\xi_2}{M^2}    \cdot  \frac{ 4  p_1\cdot   p_2  }{ |{\bm q}|^2} \nonumber\\
&&
+  \frac{\xi_1\xi_2}{M^2}    \cdot   \frac{2  (  m_1 \lambda_1     -  m_2 \lambda_2      ) }{ |{\bm q}| }
-   \frac{\xi_1\xi_2}{M^2}   ,\, |{\bm q}|>0.  \label{Vlog=-Amplog}
\ea
The last term in (\ref{Vlog=-Amplog}),
$- \frac{\xi_1\xi_2}{M^2}  $, is effective to a Feynman rule of a vertex for a four-fermion contact term, so we will drop it in the non-relativistic limit
due to the probability conservation law in the non-relativistic quantum mechanics formalism.

Then, by performing the inverse Fourier transformation
$V({\bm x})  =  {\mathcal F}^{-1}[\widetilde{V}({\bm q})] $,
we can get the potential in the coordinate space (with $|{\bm q}|>0 $ equivalent to a step function $\theta( |{\bm q}|)$, $\gamma_E$ the Euler constant) as
\ba
V(r)
&=&
  \frac{\alpha_1  \alpha_2 \Lambda^2 }{8\pi} r
-  \frac{ \Lambda ( 2 \alpha_2 \xi_1  m_1 \lambda_1     -2 \alpha_1 \xi_2   m_2 \lambda_2     )}{M}\cdot
\left[   -\frac{1}{2\pi^2 } ({\rm log\,} \frac{r}{r_0}+\gamma_E -1)   \right] \nonumber\\
&&
+   \frac{ \Lambda (\alpha_1 \xi_2   +  \alpha_2 \xi_1 )}{M}\cdot    \frac{1}{4\pi r }
-  \frac{4 \xi_1\xi_2  p_1\cdot   p_2 }{M^2}    \cdot   \frac{1}{4\pi r }       \nonumber\\
&&
+  \frac{2 \xi_1\xi_2  (  m_1 \lambda_1     -  m_2 \lambda_2      ) }{M^2}    \cdot   \frac{1}{4\pi^2 i r } \delta(r)  ,\quad (r > 0),  \label{Vr-log}
\ea
with $r_0=1\,GeV^{-1}$ put by hand to balance the dimension.
The last term of the $\delta(r)$ function in (\ref{Vr-log}) is from the $ \frac{1 }{  |{\bm q}| }$ term in (\ref{Vlog=-Amplog}), and it could also be dropped due to $r\neq 0$. At last,
with the values  $Y_1 = -1$, $Y_2 = +1$,
we have
\ba
\alpha_1\xi_2
&=& \alpha \xi  ( Q_\alpha)_{1} (Q_\xi)_{2}
= \alpha \xi  (Y \cdot 1)_{1} (Y  \cdot Y)_{2}
= - \alpha \xi , \label{alpha1-dot-xi2}\\
\alpha_2\xi_1
&=& \alpha \xi  ( Q_\alpha)_{2} (Q_\xi)_{1}
=  \alpha \xi ( Y \cdot 1 )_{2}  (Y  \cdot Y)_{1}
=  \alpha \xi ,   \label{alpha2-dot-xi1}\\
\xi_1\xi_2
&=& \xi^2   ( Q_\xi)_{1} (Q_\xi)_{2}
= \xi^2  (Y \cdot Y)_{1} (Y  \cdot Y)_{2}
= \xi^2 , \label{xi1-dot-xi2}
\ea
by combining with $\alpha_1 \alpha_2 = - \alpha^2$ in (\ref{alpha1-dot-alpha2}) and
$m_1 \lambda_1  +  m_2 \lambda_2 = 0$ in (\ref{lambda-sign}),
we can get the potential form
\ba
V(r) &=&
-   \frac{\alpha^2 \Lambda^2 }{8\pi} r
-  \frac{4  \xi^2   p_1\cdot   p_2 }{M^2}    \cdot   \frac{1}{4\pi r }     \non\\
&&
+  \frac{  \alpha  \xi \Lambda (    m_1 \lambda_1     +   m_2 \lambda_2     )}{\pi^2  M}\cdot   {\rm log\,} \frac{r}{r_0} ,
\quad  r> 0.  \label{Vr-log-0}
\ea
As expected, the linear potential also arises in (\ref{Vr-log-0}) is the same as in (\ref{U-whole-potential}), which could be corresponding to the dark energy effect (or the gravitational red-shift) and the inflation effect in a Big-bang universe. And the second term  in (\ref{Vr-log-0}) is happily  to be the Newton's gravity form!
Besides, a potential term with form of $-\frac{{\bm v} ^2 }{r}$
included in the factor
\ba
p_1 \cdot p_2 = p_1^0 p_2^0 - {\bm p}_1\cdot {\bm p}_2
\simeq \frac{ m_1}{\sqrt{1-  {\bm v}_1^2 }  } \cdot \frac{m_2}{ \sqrt{1-  {\bm v}_2^2 } } + m_1^2 |{\bm v}_1|^2
\ea
with ${\bm p}_2=-{\bm p}_1 $ in the center-of-mass frame,
can be treated as one of the source of the dark matter effects\cite{DM},
which is just of a relativistic effects!
Moreover, there will be an extra relativistic corrections from the spinor basis $u^s(p)$, by replacing $m$ to $p^0$ in (\ref{spinor-basis}), as
\be
\bar{u}^{s'}(p')u^{s}(p)=2 p^0 \delta^{ss'},\,\bar{u}^{s'}(p')\gamma^{\mu}u^{s}(p) \xlongequal{\mbox{\scriptsize(NR\,limit)}} v^{\mu}2m\delta^{ss'} \,.
\label{spinor-basis-relativity}
\ee
The logarithmic term in (\ref{Vr-log-0}) would also be treated as one of the source of the dark matter effects\cite{DM} 
in the case of $m_1 \lambda_1  +  m_2 \lambda_2 > 0$, 
which means 
some kinds of $C$ parity asymmetry effects arise in addition to (\ref{lambda-sign}), i.e., 
$\alpha_2 \xi_1  m_1 \lambda_1   \neq \alpha_1 \xi_2   m_2 \lambda_2  $;  
and, $r_0$ should be big enough so that 
the dark matter effects would be at least of the same order of the Newton's gravity.

In the sense of the superficial degree of divergence, the gravitational interaction term in (\ref{L-log}) is renormalizable, so, a construction of a renormalizable gravitation theory might be practicable in our P4 type formalism. And, this P4 formalism might also be useful to renormalize the scalar QED or the chiral perturbative theory, etc.

Besides, we want to point out that, for a $N$-body system, potential terms in (\ref{Vr-log-0}) will be additive and they will be enlarged only by the factor $(N Q_1) \cdot (NQ_2)$, rather than $(N \xi_1)(N m_1)\cdot (N \xi_2)(N m_2)$.

\section{Induced theories in some limit cases}\label{section-limit-case}

\subsection{Effects of the nonzero $\langle U\rangle$}\label{section-nonzero-VEV}

Now that $U$ is a kind of higgs field, it should show its higgs-like property.
According to the higgs mechanism, with the interaction term $\alpha  \Lambda\, \bar{\psi} U\psi $ in (\ref{L-I}), the fermion (or similarly, the boson) matter fields will get a mass correction
\be
\Delta m \thicksim \alpha \Lambda\langle U \rangle.
\label{fermion-mass-correction-U-scalar}
\ee
For a very small $\Delta m$ value, it might serve for the mass of very light particles as dark matter candidates,
or, instead of the axion\cite{axion}, it might present a solution to the strong CP (naturalness) problem. \\

If we set $\langle U \rangle_{G} = \frac{1}{L}\simeq 10^{-41} GeV$ as the gauge symmetry breaking energy scale of gravitation,
with $L\simeq 10^{11} l.y.$ corresponding to the size of the universe,
and  $\langle U \rangle_{EW} \simeq 10^2 GeV$ as the gauge symmetry breaking energy scale of electroweak interaction,
we will get a lucky coincidence for the ratio of the magnitudes of Newton's gravity force $F_{G}$ and the Coulomb force $F_{C}$,
\be
\frac{F_{G}}{F_{C}}=\left[G\left(\frac{m_{e}}{e}\right)^2 \frac{e^2}{r^2}\right]/ \left[ k  \frac{e^2}{r^2}\right] \simeq 10^{-43} \rightarrow \frac{\langle U \rangle_{G}}{\langle U \rangle_{EW}} \label{ratio-G-EW}
\ee
where $m_{e}$ is the mass of electron, $k\simeq 9\times 10^9 (N\cdot m^2 \cdot C^{-2})$ is the Coulomb constant(in SI unit).
If this is true, we might say, the smallness of gravitation constant $G$ comes from its small VEV $\langle U \rangle_{G}$ (or the huge size of the universe).

Furthermore, if we set $\langle U \rangle_{TC} $ as the gauge symmetry breaking energy scale of the technicolor (TC) interaction\cite{Weinberg}, and the ratio
\be
\frac{\langle U \rangle_{TC}}{\langle U \rangle_{EW}} \simeq \frac{g_s^2}{e^2} \simeq  \frac{0.1^2}{0.01^2} =100 \label{ratio-TC-EW}
\ee
will give us a value of $\langle U \rangle_{TC}\simeq 10^4 \, GeV =10\,TeV$ for the typical energy scale of technicolor dynamics.

\subsection{Field $U$ out a nutshell: generation of nonlinear Klein-Gordan equation}\label{U-nutshell}

Here we need the self-interaction term of $U$, which could be written as
\be
{\mathcal L}_I=  - g_U \Lambda_U^2   U \partial_{\mu} U \partial^{\mu} U  + m_U^4 U^2\,.    \label{L-int-of-U}
\ee
For a pure $U$-field system, if its kinetic energy is very small, down to $p^2 \ll \Lambda_U \Lambda$ (or, in the sense of de Broglie wavelength, we can say, the system is ``out of a nutshell"), then the kinetic energy term could be dropped, then we can get a E.O.M for $U$ according to the Euler-Lagrangian equation, as
\ba
g_U \Lambda_U^2      (\partial  U )^2  -  2 g_U \Lambda_U^2   U   \partial^2 U = m_U^4 U
\,\Rightarrow\,     (\partial  U )^2    -  2    U   \partial^2 U = \frac{m_U^4}{g_U \Lambda_U^2} U    \,.
\ea
Apparently, that is a nonlinear 2nd-order differential equations, so, we just call it ``nonlinear Klein-Gordon equation". Particularly, for a special case, $\langle U \rangle \gg U-\langle U \rangle$ (i.e., the VEV large and the fluctuation small) and $\langle U \rangle \gg \partial  U$ (i.e., the VEV large and the kinetic energy small), we can get the ``linear" Klein-Gordon equation
\be
-    \partial^2 U = \frac{m_U^4}{2  g_U  \langle U \rangle  \Lambda_U^2} U\,,   \label{deduce-linear-KG-U}
\ee
and there should be the relation $2  g_U  \langle U \rangle  \Lambda_U^2 =m_U^2$. As said for (\ref{set-U-VEV=1}),
we should remind ourselves that $\langle U \rangle $ could be very large even when the energy scale is very low!\\

In a Lagrangian, there should be both the kinetic energy terms and the potential energy terms. However, there exists the freedom to choose which ones are the kinetic energy terms and which ones are the potential energy terms, that depends the choice of the d.o.f of the system. This is a kind of ``kinetic-potential duality".

\subsection{The constraint $U_1^2 +U_2^2 = \langle U\rangle^2$ to a spontaneous breaking $U(1)$ symmetry}\label{section-Light-Constraint}

\subsubsection{$U$ as a group element: the generation of gauge field $A^\mu$}\label{U-as-group}

To a spontaneous breaking $U(1)$ symmetry, if we take the constraint
$U_1^2 +U_2^2 = \langle U\rangle^2$, there will be
\ba
U=U_1+i U_2 =\sigma(x) e^{-i\phi(x)} \rightarrow \langle U\rangle e^{-ig\phi(x)} \equiv u, \label{U-angle-term}
\ea
that is, if we choose the unitary gauge condition $\sigma =0$, $U$ will become a group element.\\

In (\ref{U-angle-term}), $U_{1}$ and $U_{2}$ are both P4 type field, and $\sigma$ and $\phi$  are also both P4 type field;
$\sigma$ is purely unphysical field (i.e.,tachyon/instanton/phantom),
while $\phi$ is physical field, as said in Sect.~\ref{higgs-U}. Is the $\phi(x)$ really a detectable field?  Mathematically to say,
$\phi$ is a phase, and we can write
\ba
U \rightarrow  u= \langle U\rangle e^{-ig\phi(x)} \rightarrow
\langle U\rangle e^{-i g \left[\phi_0(x)+\epsilon n_{\mu}A^{\mu}(x)+ \epsilon n_{\mu\nu}A^{\mu\nu}(x)+...\right]}      \,,\label{expand-phase}
\ea
that means, the P4 type $\phi$ field can be generated by many different fields rather than only one field $\phi_0(x)$.
\footnote{It is reasonable, by reminding that a Dirac spinor field could even be formally constructed as the square root of a scalar field $\phi$.}
If only the $A^{\mu}(x)$ field is nonzero in (\ref{expand-phase}),
then, with
\ba
 \bar{\psi}(U\partial\!\!\!/U^{\dag}) \psi   \rightarrow  \bar{\psi}(u\partial\!\!\!/u^{\dag}) \psi   \rightarrow  e\bar{\psi}A\!\!\!/ \psi ,
\label{dof-dU-UdU-U-scalar}
\ea
as a 4-particle-coupling term becoming to a 3-particle-coupling term, we get the gauge interaction term, with
\be
\beta \cdot\langle U\rangle^2   = e \,.
\ee
Now, instead of the d.o.f. of $\phi(x)$, there exists a connection field (gauge filed) $A^\mu(x)$, induced by the  Maurer-Cartan 1-form of $u(x)$ field.
Thus, the superficial {\bf gauge symmetry} of the Lagrangian arises!
We name the constraint
\ba
U_1^2 +U_2^2 = \langle U\rangle^2, A^{\mu}(x)\neq 0, \phi_0(x)=A^{\mu\nu}(x)=...=0 \label{Light-Constraint}
\ea
as ``{\bf Light Constraint}", in the reason that it survive only the field $A^{\mu}$ with the light speed after freezing the unphysical tachyon d.o.f. $\sigma(x)$ in (\ref{U-angle-term}) with speed over the light.

However, when both $U_{1}$ and $U_{2}$ are excited,
the contribution of the massless $U$ field
includes
an effect of a massless gauge field $A^{\mu}(x)$, see Fig. \ref{fig-tree-LO}-(a).
Now, as both the $\bar{\psi} \partial U   \psi $
term and the $\bar{\psi}A^\mu \psi$ term can generate the Coulomb potential,
we would like to ask, is the gauge symmetry necessary? We will return this question in Sect.~ \ref{section-causality}.

\subsubsection{Multi-vacuum structure for sine-Gordon type vector field $A^{\mu}$}\label{sine-Gordon-A}

\noindent 1. Multi-vacuum structure for $A^{\mu}$\\

If we write
\footnote{As said for (\ref{set-U-VEV=1}),
we should remind ourselves that $\langle U \rangle $ could be very large even when the energy scale is very low!}
\be
U(x) = \exp[-i g \epsilon n^{\mu}A_{\mu}(x)] =\cos[g \epsilon n^{\mu}A_{\mu}(x)]-i\sin[g \epsilon n^{\mu}A_{\mu}(x)] ,\
\ee
then the potential term
\be
V(A)\thicksim U(A)+U^{\dag}(A)=\cos[(g\epsilon)\cdot A], \label{define-cosA}
\ee
would mean that the dynamics for the field $A^{\mu}$ is of a sine-Gordon type (or, a kind of generalized higgs type vector), see Fig. \ref{U-potentails}-(2).
Thus,
there might be many excitations for $A^\mu$ at different vacuums (or, VEVs), with heavy masses in the large $g$ cases( $g\epsilon \simeq 1$) and small masses in the small $g$ cases.\\

\noindent  2. Mass spectrum with generation structure \\

Like the mass correction in (\ref{fermion-mass-correction-U-scalar}) from $U$, with the term $\bar{\psi}A\!\!\!/\psi$, the fermion (or similarly, the boson) fields can get a mass correction from $A^{\mu}$,
\be
\Delta m \thicksim \alpha \Lambda\langle A \rangle \thicksim \alpha \Lambda \frac{(2n+1)\pi}{g\epsilon}   \,\,,\,\,n=0,1,2,... \,.
\label{fermion-mass-correction-A}
\ee
where the number $n$ might lead the fermion mass spectrum to a generation structure. Even for the same value of $n$, we can get the deductions below:\\
\indent a. if $\Delta m$ is the mass differences between the current quarks and the constituent quarks, then, by setting
\be
g \thicksim  \frac{(2n+1)  \alpha \Lambda}{\Delta m \cdot\epsilon} \xlongrightarrow{{\mathcal O}(\Lambda)\thicksim {\mathcal O}(\epsilon)}  \frac{(2n+1)  \alpha }{\Delta m } \thicksim 1 \,,
\ee
with $\Delta m \thicksim 1GeV$ and $n=0$, we have $\alpha \thicksim 1$.\\
\indent b. if $g \thicksim  0.01$ for the E.W. interaction, then, $\Delta m \thicksim 100GeV$, corresponding to the possible heavy fermions.\\

\noindent 3. A seesaw mechanism for gauge symmetry and flavor symmetry \\

See Fig. \ref{U-potentails}-(2), with (\ref{define-cosA}), for a vacuum at $A=\langle A\rangle_i$, the potential could be written as
\be
V(A\simeq A_i)\simeq  -1 + (g\epsilon)^2 (A-A_i)+\ldots,
\ee
which means the mass of the excitation $A'=A-\langle A\rangle_i$ is of order $\thicksim  m= g\epsilon $. So, we can get the conclusions below:\\
\noindent (1) when $g\rightarrow 0$,\\
\indent a. $A'_{\mu}$ is nearly massless, so the gauge symmetry is restored;\\
\indent b. the VEV $\langle A\rangle_i$ are of very different magnitudes, so, through (\ref{fermion-mass-correction-A}), the fermion masses would be also of very different magnitudes, including very heavy fermions; this is a kind of flavor symmetry breaking for fermions;\\
\noindent (2) when $g\rightarrow \infty$,\\
\indent a. $A'_{\mu}$ is massive, with the diagonal elements in its mass matrix being large, so the gauge symmetry is broken;\\
\indent b. since the unphysical d.o.f (i.e.,tachyon/instanton/phantom) $\sigma$ in (\ref{U-angle-term}) was excited now, the vacuum tunnelling (oscillating) effect would become strong, so the off-diagonal elements in the mass matrix of $A'_i$ become large, too; or, in another viewpoint, now it's $A'_{\mu}$ that was frozen, and the tachyon was the real d.o.f for mediating interactions; we can treat the tachyon massless or nearly massless according to the absence of heavy bosons in a hadron;\\
\indent c. the VEV $\langle A\rangle_i$ in the neighbour minimum are nearly equal, so, there would be a degenerate for the fermion mass, or, we can say, the flavor symmetry for fermions would be restored; besides, it's now allowed for very small fermion masses through (\ref{fermion-mass-correction-A}), which might be an underlying reason for the feasibility of the  ``large $N_c$" or ``large $N_f$" hypothesis for a real hadron, and for the possible neutrino
 oscillation.

So, maybe this is a new kind of dynamical symmetry breaking/restoring mechanism, with a seesaw for gauge symmetry and flavor symmetry. \\

\subsubsection{Duality between matter fields and media fields: from non-perturbative to perturbative}\label{duality-matter-media}

\noindent {\bf 1. Matter fields are P2 type, while media fields are underlying P4 type.}\\ 

Instead of the gauge field $A^\mu \thicksim u \partial u^\dag $ ($u$ is a group element), the employment of the Wilson line $U(y,x)$ and Wilson loop $U_P(x,x)$, which are defined as\cite{Peskin}
\ba
U_P(x+\epsilon n,x)  &=& 1-i g \epsilon n^{\mu}A_{\mu}(x) +  {\mathcal O}((g\epsilon)^2) ,\label{U-definition}\\
U_{P_{ij}}(x,x)&=& 1 -i \epsilon^2 g  F_{ij}  + {\mathcal O}(\epsilon^3),\label{plane-expansion-ij}
\ea
ensured the availability of lattice gauge theory.
It is just this subtle hint that inspired us to consider a field $U$,
with a hidden correspondence of  the Wilson loop $U_P$,
\be
U_P\rightarrow U\,,
\ee
rather than the gauge field $A$
as a possible effective d.o.f., with the Light Constraint in (\ref{Light-Constraint})
\ba
g A_{\mu} = u(x)i\partial_{\mu}u^{\dag}(x) \rightarrow  U(x)i\partial_{\mu}U^{\dag}(x), .
\ea
Thus, as an inverse procedure, it is a useful try to solve the non-perturbative problem in stong QED by defining a P4 type complex scalar field $U=U_1+iU_2$ with $U_1$ and $U_2$ are both excited, instead of the group element $u$.\\

\noindent {\bf 2. Media fields are P2 type, while matter fields are underlying P4 type.}\\ 

Besides the media field $A$, we can also treat the fermion matter field $\psi$ as P4 type field. For convenience, we choose a scalar matter field $\phi$ and take the scalar QED as an example to illustrate our motivation.

If we treat the field $\phi$ as effective reduction of underlying P4 type field $\Phi$, then the P2 type current of $\phi$ will become a P2 type field, as
\ba
J^{\mu}(x) = \phi^{\dag}i\partial^{\mu} \phi(x)
&\rightarrow &
\Phi^{\dag}i\partial^{\mu} \Phi(x)\equiv \mathcal{A}^{\mu}(x),\non\\
\mbox{(P2 type field $\phi$}  &\rightarrow& \mbox{ P4 type field $\Phi$)}\rightarrow \mbox{(Maurer-Cartan 1-from of $\Phi$)}, \non\\
\mbox{(P2 type current $J^{\mu}$)} &\rightarrow&\mbox{(P2 type field)}   .
\label{current-to-field-eq}
\ea
It is reasonable for (\ref{current-to-field-eq}), since the only difference between a current and a vector field is that: a field has a E.O.M, while a current hasn't; for other things, they could be treated as the same.

Thus, with the Light Constraint in (\ref{Light-Constraint}), the old  P2 type (nonrenormalizable) 3-particle interaction term  will become a new 2-particle mixing term (which will be a perturbative ``interaction term"), as
\ba
{\mathcal L}_{I}  =   e   A_{\mu}  \phi i \overleftrightarrow{\partial}^{\mu}\phi
\rightarrow
e   A_{\mu}  \Phi i \overleftrightarrow{\partial}^{\mu}\Phi = e   A_{\mu}   \mathcal{A}^{\mu} ={\mathcal L}_{K}.
\ea

Besides, it seems like that the new d.o.f. $\mathcal{A}^{\mu}$ in the limit of $\phi \rightarrow \Phi$ could propagate to a composite system of two collinear $\phi$, so,
the generation of the new mixing term (or, the new d.o.f. $\mathcal{A}^{\mu}$) is associate with collinear motion of the two $\phi$ particle, which is also a kind of ``kinetic-potential duality".\\

\noindent {\bf 3. On the generalization of a theory}\\

On the generalization of a theory, one method might be to {\bf extend} the d.o.f, such as, to introduce greater symmetry, more particles and more interactions, more extra dimensions, or more complicate rules, mathematically, by introducing more complicate groups,  more complicate variables (e.g., complex, quaternion or octonion valued), more coordinates, higher-order and nonlinear equations, etc.
If the results in our calculations are useful for the real physical processes, then it would be said that the P4 type theory is a more general theory than the P2 type ones.

Another method might be to {\bf redefine} the effective d.o.f for a system, such as: the Bogoliubov transformation; the wave-particle duality, in which
the redefinition of the canonic d.o.f from the momentum (current) $p$ to the wave ($p \rightarrow \phi$) for the first quantization in quantum mechanics;
or, from the P4 type current to the P2 type field ($J =\Phi i\partial \Phi  \rightarrow {\mathcal A}$) in this paper. According to these examples, maybe we could ask, is there a principle about this redefinition of d.o.f, or maybe we can call it ``materialization" (from variable to matter)? 

Besides, when the dynamics of a variable (or, d.o.f) becomes complicate, maybe it is the time to redefine the kinetics rather than the interactions for the systems, such as the chaos or the turbulence systems, or non-perturbative or the nonrenormalizable systems.
Or, maybe we can give a hypothesis that, all the good (or, well-defined) theories should be simple (or, perturbative, linear), while all the complicate (or, non-perturbative, nonlinear) theories should be due to the bad choice of the d.o.f.

\section{The causality in our P4 theory} \label{section-causality}

\subsection{The canonic commutator}

Firstly, let's solve the E.O.M in (\ref{EOM-U}).
For simplicity, we will only consider the $U_1$ component of $U=U_1+iU_2$ in the case of $\Lambda_{U}=0$, i.e.,
\ba
-\partial^{\mu}\partial^{\nu}\partial_{\mu}\partial_{\nu}U_1   =-m_{U}^4 U_1  ,\,p^4=m_{U}^4.  \label{EOM-of-U1}
\ea
Then, by combining the plane-wave solution (\ref{field-solution-of-plane-wave}) and the instanton solution (\ref{field-solution-of-instanton}),
we might get the general solution with the form of
\ba
U_1(x) &=& c_1 e^{ip\cdot x}+c_2 e^{-ip\cdot x} +
d_1  e^{p\cdot x}+d_2 e^{-p\cdot x},
\label{general-solution-U1-0}
\ea
however, we will write the general solution with the form of
\ba
U_1(x ) =  \alpha         a_{\bm p}         e^{-i p\cdot x}  +  \alpha^{\dag}  a_{\bm p}^{\dag}  e^{ i p\cdot x}, \label{general-solution-U1}
\ea
that is to say, the effects of the unphysical
solution are absorbed into the coefficients $\alpha$ and $\alpha^{\dag}$.\\

Secondly,
we will introduce a new {\bf postulation} for the canonic quantization of our P4 type field $U$.
By taking the $U_1$ component of $U=U_1+iU_2$ as an example,
we can express the canonic quantization results as below:\\
(1) canonic variables of field $U_1$ (with Eq. (\ref{general-solution-U1})):
\ba
U_1(x ) &\equiv& \int \frac{d^3 p}{(2 \pi)^3}
\frac{1}{p^0  \sqrt{2  p^0}}
\left(
 \alpha_{\bm p}         a_{\bm p}         e^{-i p\cdot x}
+\alpha_{\bm p}^{\dag}  a_{\bm p}^{\dag}  e^{ i p\cdot x}
\right),\\
\Pi_1(x )\equiv  \partial_t \partial^2 U_1(x )   &=& \int \frac{d^3 p}{(2 \pi)^3}
\frac{i (p\cdot p)}{\sqrt{2  p^0}}
\left(
    \alpha_{\bm p}      a_{\bm p}         e^{-i p\cdot x}
-   \alpha_{\bm p}^\dag a_{\bm p}^{\dag}  e^{ i p\cdot x}
\right) ;
\label{canonic-variables-of-U1}
\ea
(2) the creation and annihilation operators, state vectors and normalization relations:
\ba
&&  \bar{{\bm p}}^2  \equiv(p^0)^2 \pm   m^2, \quad \mbox{with}\quad  p^2 \equiv p\cdot p =  \pm   m^2; \\
&&  \alpha |0\rangle \equiv  0,\, |0^{\ast}\rangle \equiv \sqrt{\frac{ (p\cdot p)   \bar{{\bm p}}^2}{(p^0)^4}  } \alpha_{\bm p}^\dag |0\rangle,\,\non\\
&&  \langle 0 |0\rangle =  \langle 0^{\ast}| 0^{\ast}\rangle  =1 ;\,\non\\
&&  a |0\rangle \equiv  0,\,
|{\bm p}\rangle  \equiv  \sqrt{  2  p^0   }  a_{\bm p}^\dag   |0\rangle, \, 
|{\bm p}^{\ast}\rangle  \equiv   \sqrt{  2  p^0   }   a_{\bm p}^\dag   |0^{\ast}\rangle, \, \non\\
&&
\langle {\bm p}|{\bm q}\rangle = \langle {\bm p}^{\ast}|{\bm q}^{\ast}\rangle = 2  p^0   (2 \pi)^3 \delta^{(3)}({\bm p} -{\bm q} )   ; \label{state-of-U1}\\
&&
\langle 0| U_1(x) |{\bm p}^{\ast} \rangle =  \frac{1}{p^0}    e^{-i p\cdot x} ;\,
\ea
that means, we define the operator $\alpha_{\bm p}$ or $\alpha_{\bm p}^\dag$ with the operation of shifting 
a stable physical vacuum $|0\rangle$ to another unstable unphysical vacuum $|0^{\ast}\rangle$, 
and the effect of field operator $U_1$ is to annihilate an unstable state $|{\bm p}^{\ast} \rangle$ created from an arbitrary unstable vacuum $|0^{\ast}\rangle$; 
and, in the limit case that $\alpha_{\bm p}$ and $\alpha_{\bm p}^\dag$ are $c$-numbers, 
only the physical vacuum $|0\rangle$ and the physical states $|{\bm p} \rangle$ exist;  \\
(3) canonic commutators:
\ba
\left[U_1 ({\bm x},t ),  \Pi_1({\bm x}' ,t)\right]
=\left[U_1 ({\bm x} ,t), \partial_t \partial^2 U_1 ({\bm x}' ,t) \right]
&=& - i \delta^{(3)}({\bm x} -{\bm x}' ) \cdot\frac{(p^0)^2}{   \bar{{\bm p}}^2} ,
\non\\
\,\Leftrightarrow\,
\left[\alpha_{\bm p}  ,\alpha_{{\bm p}'}^\dag \right] =\frac{(p^0)^4}{ (p\cdot p)   \bar{{\bm p}}^2}, \,
\left[ a_{\bm p} ,  a_{{\bm p}'}^\dag \right] &=&  (2 \pi)^3 \delta^{(3)}({\bm p} -{\bm p}' ), \non\\
\left[\alpha_{\bm p} a_{\bm p} ,\alpha_{{\bm p}'}^\dag a_{{\bm p}'}^\dag  \right] 
&=& (2 \pi)^3 \delta^{(3)}({\bm p} -{\bm p}' ) \frac{(p^0)^4}{ (p\cdot p)   \bar{{\bm p}}^2} , \non\\
others &=& 0; \label{commutator-of-U1}
\ea
(4) Legendre transform and Hamiltonian:\\
as the Lagrangian can be rewritten to be
\ba
{\mathcal L}_{U_1} &=& -  \partial_\mu  \partial_\nu U_1({\bm x},t)  \partial^\mu  \partial^\nu  U_1({\bm x},t)  + m_U^4 [U_1({\bm x},t)]^2 \non\\
&=&\partial_\nu U_1({\bm x},t)     \partial^\nu [\partial^2 U_1({\bm x},t) ]
-\partial_\mu  [\partial_\nu U_1({\bm x},t)  \partial^\mu  \partial^\nu  U_1({\bm x},t)  ]
+ m_U^4 [U_1({\bm x},t)]^2 \non\\
&=& \dot{U}_1({\bm x},t)     \partial_t [\partial^2 U_1({\bm x},t)] - \nabla U_1({\bm x},t) \cdot \nabla[\partial^2 U_1({\bm x},t)]
+ m_U^4 [U_1({\bm x},t)]^2 ,\label{surface-term}
\ea
where the second
term (the total derivative) in the second line in (\ref{surface-term}) can be dropped because the corresponding surface term is zero on the boundary of the space, the Hamiltonian can be get from the Legendre transform
\ba
H_{U_1} &=& - \int d^3{\bm x}  \left[ \Pi_1({\bm x},t) \dot{U}_1({\bm x},t)-  {\mathcal L}_{U_1}({\bm x},t) \right]  \non\\
&=& +\int \frac{d^3 p}{(2 \pi)^3}
\left[\frac{    (p\cdot p)   \bar{{\bm p}}^2  }{    (p^0)^3 }  \right]
\frac{1}{2}\left( \alpha_{\bm p}^{\dag} a_{\bm p}^{\dag} \alpha_{\bm p} a_{\bm p}
+  \alpha_{\bm p}  a_{\bm p}   \alpha_{\bm p}^{\dag} a_{\bm p}^{\dag}  \right)
\non\\
&=& +\int \frac{d^3 p}{(2 \pi)^3}
p^0 
\frac{1}{2}\left(  a_{\bm p}^{\dag}   a_{\bm p}
+   a_{\bm p}     a_{\bm p}^{\dag}  \right)   , \label{Hamiltonian-of-U1}
\ea
and the Schrodinger equation can be get as
\ba
H_{U_1} |{\bm p}\rangle  = p^0 |{\bm p}\rangle  .
\ea

Thirdly, we want to emphasize that,
although
there are fourth-order derivative terms and unphysical solutions in our P4 type field theory,
however,
(1) according to the Hamiltonian in (\ref{Hamiltonian-of-U1}), $H\sim {\bm p}^4 \neq (\dot{{\bm p}})^2 \neq  \partial_t^4 {\bm x}$,
the corresponding classic dynamics will include only 2nd-order derivative terms (with ${\bm p}^2 {\bm p} \sim {\bm p}^3$ as ``cononical momentum" and ${\bm p}^4 \sim ({\bm p}^3)^2/{\bm p}^2$ in the Hamiltonian to get an E.O.M of the form ${\bm p}^2 \ddot{\bm x}+ \kappa {\bm x}=0$)
so that there will not be acausality;
(2) only two canonical variables, $U$ and $\partial_t \partial^2{U}$, are enough to construct our theory,
that is to say, it is not necessary to define multiple canonic variables, e.g., extra extended conjugate momenta such as $\dot{U}$ and $\ddot{U}$,
although they are needed as initial conditions to fix the 4 coefficients in (\ref{general-solution-U1-0}) or (\ref{general-solution-U1}) 
in the classic mechanics.\\

\subsection{The causality defined by a generalized propagator}

Let's go back to the causality topic mentioned
in Section \ref{ddU-in-LK}.
Here are the different expressions to causality in classic mechanics and quantum mechanics:\\
\indent (a) in classic mechanics, the causality depends on the interval of the variables in the coordinate space, i.e., whether the interval is time-like or not;\\
\indent (b) in the {\bf Heisenberg picture} for quantum mechanics,
the causality  depends on the ``interval" (defined by the commutator) of the variables (operators) in the algebra space, i.e., whether the ``interval" is time-like or not;\\
\indent (c) in the  {\bf path integral} formalism, the causality depends on the time-order operator $\hat{\rm T}$ inserted for the Feynman propagators due to the retard potential boundary condition; etc.\\

We want to propose that unphysical d.o.f does not mean acausality. In our P4 type field theory formalism, the causality is rigid, since the correlation function defined in (\ref{define-correlation-function}) is also expressed with the time-order operator $\hat{\rm T}$! What we need to do
is only to interpret the effects of the unphysical d.o.f.

Here we rewrite the correlation function in (\ref{define-correlation-function}) as
\be
D_F(x-y) \equiv  \langle \Omega_2| \hat{\rm T} U(x)U(y)|\Omega_1\rangle ,  \label{define-correlation-function-2}
\ee
where $|\Omega_1\rangle$ and $|\Omega_2\rangle$ are two vacuum states.
If we write the ``interval" between two vacuum states $|\Omega_1\rangle$ and $|\Omega_2\rangle$ as their inner product $\langle\Omega_1|\Omega_2\rangle \equiv e^{i \theta}$, the value of $\theta$ can be real or complex. A real-valued  $\theta$ is associated with the vacuum tunnelling processes among the so-called stable ``$\theta$ vacuum", and the complex-valued $\theta$  would be associated with the more general  vacuum evolution dynamics. Our postulation is that, effects from the unphysical d.o.f of $U(x)$ and the non-unitary vacuum transition processes are combined to a physical propagator.

The unphysical things are indeed physical.
Some examples are listed as below:\\
\indent (1) In the ``unphysical limit", i.e., $V(U)=-  m_{U}^4 U^{\dag}U =
+ \lambda_{U} \Lambda_{U}^4 U^{\dag}U U^{\dag}U=0$ in (\ref{potential-U}), the fields $U_1$ and $U_2$ (or, $\sigma(x)$ and $\phi(x)$) in (\ref{U-angle-term}) are both excited, and now, the higgs/tachyon/instanton/phantom effects are excited completely, which will be reflected in the detectable world. Now we might understand the global $U(1)$ symmetry of $U$ field as a kind of symmetry between the inner region and the outer region of the light cone.\\
\indent (2) In the ``physical limit", or the``Light constraint", $U_1^2+U_2^2=\langle U\rangle^2$, see (\ref{Light-Constraint}) in Sect.~\ref{U-as-group}, that is, in the vacuum symmetry breaking case,
the gauge symmetry would arise automatically; the vacuum states are all stable, and, only the physical d.o.f $e^{-ip\cdot x}$ and speed value $c=1$ for the light survive as $t\rightarrow \infty$, so their effects can be physical and detectable all the time.\\
\indent (3) In the ``meta-physical case", if the ``Light constraint" is not rigidly satisfied, then the unphysical partner of light would exist and the speed of light would fluctuate; although the so-called unphysical d.o.f $e^{-p\cdot x}$ in (\ref{EOM-tachyon},\ref{EOM-phantom}) are unstable, their residual effect should be detectable until $t\rightarrow \infty$ (no matter the momentum $p$ of field $U$ is large or small). In other words, nontrivial vacuum could be treated as potential barrier background, so an attenuation (imaginary-valued momentum) is normal for a particle transit through the potential barrier.  \\

By combining the interpretation to causality above, maybe we can introduce a new terminology called ``{\bf propagator picture}", that is, we interpret the causality by introducing a generalized path integral formalism, including the unphysical particle d.o.f and the unphysical vacuum but generating physically causal amplitudes.
In a word, the inclusion of the unphysical particle d.o.f and vacuum evolution dynamics is the origin of the differences between our P4 type field theory and the P2 type theories.

\section{Conclusion}\label{conclusion}

We have introduced a new class of higgs type complex-valued scalar fields $U$ (``P4 type")  with a fourth-order differential equation as its equation of motion, motivated by the linear potential in the lattice gauge theory, and we have seen something new in a theory which can generate a linear potential on the level of effective theories.
The field $U$ can generate a wealth of interaction forms with some postulations on the convergence being taken.
After getting a propagator of the form of $-i/p^4$ from a $(\partial \partial U)^2$ term in the kinetics term in the canonic quantization formalism,
by computing the amplitudes
of the tree-level  $2\rightarrow 2$  scattering processes mediated by the $U$ field,
we can get a wealth of classic non-relativistic effective potential form within the Born-approximation formalism, such as:
(1) by using $U$ to construct a QED theory, we can get the Coulomb-type potential, with a negligible linear potential and logarithmic potential as correction;
(2) by using $U$ to construct a QCD theory, we can get the Coulomb-type potential, and a considerable linear potential to serve for the confinement,
with a logarithmic potential as the next-leading order corrections;
(3) by using $U$ to construct a gravatition theory, we can get 
a linear potential to serve for the dark energy effect and the inflation effect, 
a Coulomb-like potential to serve for the Newton's gravity, 
and a logarithmic potential 
combining with a relativistic correction to Newton's gravity 
to serve for the dark matter effect; 
in the sense of the superficial degree of divergence, this gravitation theory is renormalizable, so, a construction of a renormalizable gravitation theory might be practicable in our P4 type formalism.

Moreover, in some limit cases, we can get some interesting deductions, such as:
(1) in a low energy approximation of the dynamics of $U$, a nonlinear Klein-Gordon equation could be generated;
(2) with a constraint $U_1^2 +U_2^2 = \langle U\rangle^2$ to a spontaneous breaking $U(1)$ symmetry, $U$ could become a group element, thus the gauge symmetry could superficially arise, with a linear QED to be generated by relating the field strength  $\partial U$ to the corresponding gauge field $A^\mu$;
(3) due to the multi-vacuum structure for a sine-Gordon type vector field  $A^\mu$ induced from $U$, a mass spectrum with generation structure and a seesaw mechanism on gauge symmetry and flavor symmetry could be generated, including heavy particles;
(4) by treating the P2 type matter fields as the effective d.o.f of P4 type ones (with a kind of ``kinetic-potential duality"),
or, by treating the P2 type gauge field $A^\mu$ as the effective d.o.f of P4 type $U$ fields (with a correspondence to the Wilson line $U_P$),
it provides a possible way to deal with the non-perturbative problems.
So, a solution to the non-perturbative problems might be practicable in our P4 type formalism.

For the causality,
we interpret the causality by introducing a generalized path integral formalism, including the unphysical particle d.o.f and the unphysical vacuum but generating physically causal amplitudes. In a word, the inclusion of the unphysical particle d.o.f and vacuum evolution dynamics is the origin of the differences between our P4 type field theory and the P2 type theories.


\section{Acknowledgements}

The author is very grateful to Prof. Xin-Heng GUO at Beijing Normal University, Dr. Xing-Hua WU at Yulin Normal University and Dr. Jia-Jun Wu at University of Chinese Academy of Sciences, for very important helps.



\begin{thebibliography}{}
\addcontentsline{toc}{section}{References}


\bibitem{HuangCY-book}
C.Y. Wong, ``{\it Introduction to High-Energy Heavy-Ion Collisions}", World Scientific Publishing, (1994). Chapter 10-11; Page 207,214; Eq.(10.63),(11.13).

\bibitem{Peskin}
M.E. Peskin, D.V. Schroeder, ``{\it An Introduction to Quantum Field Theory}", (Boulder: Westview, 1995). Page: 18,30,43,51,121-126,186,193,294,310,482. \\
A.Zee, ``{\it Quantum Field Theory in a Nutshell}", (Princeton University Press, 2010). Page: 518.

\bibitem{LiRC-P4-theory}
R.-C. Li, Field Theory with Fourth-order Differential Equations,\\
\url{https://vixra.org/abs/1712.0487}



\bibitem{Pais-Uhlenbeck}
A. Pais and G. E. Uhlenbeck, ``On Field Theories with Non-Localized Action", Phys. Rev. 79, 145 (1950).


\bibitem{JPHsu}
J.-P. Hsu, ``Yang-Mills Gravity in Flat Space-time, II. Gravitational Radiations and Lee-Yang Force for Accelerated Cosmic Expansion", Int.J.Mod.Phys.{\bf A24}, 5217 (2009). arXiv:1005.3250v1 [gr-qc].

\bibitem{tachyon}
G. Feinberg, ``Possibility of Faster-Than-Light Particles", Physical Review.{\bf 159}, 1089 (1967).


\bibitem{phantom}
R. R. Caldwell, Phys.Lett.{\bf B545}, 23 (2002). [arXiv:astro-ph/9908168]. \\
Y.-F. Cai, E.N. Saridakis, M.R. Setare, et al., Phys.Rept.{\bf 493}, 1 (2010), arXiv:0909.2776v2 [hep-th].


\bibitem{Weinberg}
S. Weinberg, ``{\it The Quantum Theory of Fields, Vol. I}", (Cambridge University Press, 1995). Page: 58;
ditto, Vol. II , 1996, Pages: 326; 464-468.



\bibitem{Itzykson-Zuber}
C. Itzykson, J.B. Zuber, ``{\it Quantum Field Theory}", (New York: McGraw-Hill, 1980). Page: 42-44.



\bibitem{ChengTP-book}
T.P. Cheng, L.F Li, ``{\it Gauge Theory of Elementary Particle Physics-Problems and Solutions}",(Oxford University Press, 2000). Page: 111.



\bibitem{e-measurement}
J.K. Webb, J. A. King, M. T. Murphy, et al., ``Indications of a spatial variation of the fine structure constant", Phys. Rev. Lett., {\bf 107}, 191101 (2011). arXiv:1008.3907[astro-ph.CO].


\bibitem{DM}
S. M. Faber, R. E. Jackson, ``Velocity dispersions and mass-to-light ratios for elliptical galaxies", Astrophys.J. {\bf 204}, 668 (1976).



\bibitem{axion}
R. D. Peccei, Helen R. Quinn, ``CP Conservation in the Presence of Pseudoparticles", Phys. Rev. Lett. {\bf 38}, 1440 (1977).




\end{thebibliography}
\end{document}